\documentclass[12pt]{article}

\usepackage{amssymb,amsthm,amsfonts,psfig,epsfig}

\def\gappeq{\mathrel{\rlap {\raise.5ex\hbox{$>$}} {\lower.5ex\hbox{$\sim$}}}}
\def\lappeq{\mathrel{\rlap{\raise.5ex\hbox{$<$}} {\lower.5ex\hbox{$\sim$}}}}

\def\beq{\begin{equation}} \def\eeq{\end{equation}} 
\def\bea{\begin{eqnarray}} \def\eea{\end{eqnarray}}
\def\bq{\begin{quote}} \def\eq{\end{quote}}

\def\nn{\nonumber}

\def\ti{\tilde}
\def\d{\delta}

 \def\ie{{i.e.}} \def\eg{{\it e.g.~}}

\def\m02{m_0^2}   \def\A02{a_0^2}
\def\i3{\mathbb{I}}
\def\3{N_3}  \def\2{N_2} \def\1{N_1} \def\a{N_a}  \def\b{N_b}
\def\tgb{\tan \beta} 
\def\tmg{\tau \rightarrow \mu \gamma} \def\meg{\mu \rightarrow e \gamma}

\begin{document} 

\pagestyle{empty} 
\begin{flushright}  SACLAY-T03/053 \end{flushright}  
\vskip 2cm    
\def\thefootnote{\fnsymbol{footnote}}
\begin{center}  
{\Large \bf
Lepton Electric Dipole Moments \\   \vskip .2cm  from Heavy States Yukawa Couplings 
} \vspace*{5mm}    \end{center}  
\vspace*{5mm} \noindent  \vskip 0.5cm  
\centerline{\large \bf Isabella Masina 
}
\vskip 1cm \centerline{\em Service de Physique Th\'eorique 
\protect\footnote{Laboratoire de la Direction des Sciences de la Mati\`ere du 
Commissariat \`a l'\'Energie
Atomique et Unit\'e de Recherche Associ\'ee au CNRS (URA 2306).} , CEA-Saclay}
\centerline{\em F-91191 Gif-sur-Yvette, France} \vskip2cm   
\centerline{\bf Abstract}  

In supersymmetric theories the radiative corrections due to heavy states 
could leave their footprints in the flavour structure of the supersymmetry breaking masses.
We investigate whether present and future searches for the muon and electron EDMs
could be sensitive to the CP violation and flavour misalignment induced on slepton masses 
by the radiative corrections due to the right-handed neutrinos of the seesaw model and to
the heavy Higgs triplets of $SU(5)$ GUT. 
When this is the case, limits on the relevant combination of neutrino Yukawa couplings are obtained. 
Explicit analytical expressions are provided which accounts for the dependencies
on the supersymmetric mass parameters.

\vspace*{1cm} \vskip .3cm  \vfill   \eject  
\newpage  
\setcounter{page}{1} 
\pagestyle{plain}

\def\thefootnote{\arabic{footnote}}
\setcounter{footnote}{0}


\section{Introduction}

In low energy supersymmetric extensions of the Standard Model (SM),
unless sparticle masses are considerably increased,
present limits on flavour violating (FV) decays and electric dipole moments (EDMs) 
respectively allow for a quite small amount of fermion-sfermion misalignment in the flavour basis 
and constrain the phases in the diagonal elements of sfermion masses, 
involving the parameters $\mu$ and $A$, to be rather small.
The bounds on the supersymmetric contribution to lepton (L)FV decays and EDMs 
have the advantage, as compared to the corresponding squark sector ones, 
of being not biased by the SM contribution - 
nor by the non-supersymmetric seesaw \cite{seesaw} contribution \cite{pet}. 
Experimental limits on LFV decays and EDMs are then a direct probe of the flavour 
and CP pattern of slepton masses - see e.g. Ref. \cite{ms1} for a recent collection.
From a theoretical perspective, understanding why CP phases and deviations from alignment
are so strongly suppressed is one of the major problems of low energy supersymmetry, 
the {\it CP and flavour problem}. 
On the other hand, precisely because FV decays and EDMs provide strong constraints, they can share some 
light on the features of the (possibly) supersymmetric extension of the SM.   

Indeed, it is well known that even if these CP phases and misalignments were absent - or suppressed enough -
from the effective broken supersymmetric theory defined at $M_{Pl}$, 
they would be generated at low energy by the RGE corrections due to other flavour 
and CP violating sources already present in the theory, in particular from the Yukawa couplings 
of heavy states like the right-handed neutrinos of the seesaw model \cite{bormas} and the Higgs triplets of $SU(5)$ 
grand unified theories (GUT) \cite{barb}.
Effects of radiative origin have the nice features of being naturally small, exactly calculable 
once specified a certain theory and - most interestingly - if experiments are sensitive to them, 
they yield limits on some combination of Yukawa couplings \footnote{Of course, several effects of different origin 
could be simultaneously present, but there would be no reason for a destructive interference among them to occur.}.

Recently, it has been stressed  that the present (planned) limit 
on $\meg$ \cite{megexpp} ($\tmg$ \cite{tmgexpf}) is sensitive to the lepton-slepton misalignments induced by the 
radiative corrections in the framework of the supersymmetric seesaw model and that the associated constraints
on neutrino Yukawa couplings have a remarkable feedback on neutrino mass model building
\cite{lfv_ss,  imsusy02}. 
However, LFV decays cannot provide any information on the pattern of CP violation of slepton masses, 
while EDMs are sensitive to both LF and CP violations. 
Since the present sensitivities to the electron and muon EDMs, 
$d_e < 10^{-27}$ e cm \cite{deexpp} and $d_\mu < 10^{-18}$ e cm \cite{dmuexpp}, 
could be lowered by planned experiments by up to three to five \cite{deexpf, lam} and six to eight \cite{dmuexpf,
dmuexpff} orders of magnitude respectively, 
it is natural to wonder {\it whether lepton EDMs would explore the range associated to LF and CP violations of 
radiative origin}.

In this work we analyze the predicted range for $d_e$ and $d_\mu$
when the radiative corrections due to the right-handed neutrinos of the seesaw model and, possibly, 
the Higgs triplets of $SU(5)$ GUT provide the main source of CP violation and misalignment in slepton masses. 
This allows to extract many informations. 
If experimental searches could explore this range, one could obtain limits on the 
imaginary part of the relevant combination of neutrino Yukawa couplings.
Moreover, the eventual discovery of a lepton EDM in this range might be interpreted as indirectly suggesting 
the existence of such a kind of fundamental particles which are too heavy to be more manifest.
On the other hand, finding $d_e$ and/or $d_\mu$ above this predicted range would prove the existence of 
a source of CP (and likely also LF) violation other than these heavy states. 

The pure seesaw case has been considered in Ref. \cite{ellis-1} 
- see also the related studies \cite{ellis-2, ellis-n} on specific seesaw textures -, 
where it has been pointed out that threshold effects due to hierarchical
right-handed neutrino masses enhance the radiatively-induced ${\cal I}m(A_{ii})$, $i=e,\mu,\tau$. 
By extensively reappraising this framework, we find that for $\tgb \gtrsim 10$
the amplitude with a LL RR double insertion - proportional to $\tan^3 \beta$ - 
dominates over the one involving ${\cal I}m(A_{ii})$ - insensitive 
\footnote{In Ref. \cite{ellis-2} a dependence on $\tgb$ arises due to the particular class of seesaw textures studied.} 
to $\tgb$ -, 
the exact ratio depending on the particular choice of supersymmetric mass parameters. 
We provide general expressions from which it is easy to recognize the model dependencies
and which complete previous analyses. 
Lepton EDMs are then strongly enhanced in models with large $\tan \beta$.
To see how close experiments are getting to the seesaw induced lepton EDMs range,
we compare the upper estimates for the radiatively-induced misalignments and CP phases
with the corresponding present and planned experimental limits collected in \cite{ms1}.

The range for a seesaw-induced $d_\mu$ turns out to be quite far from present searches and,
in particular, its eventual discovery above $10^{-23}$ e cm would prove the existence of a source 
of CP violation other than the neutrino Yukawa couplings. 
On the contrary, the present sensitivity to $d_e$ explores the seesaw-induced range  
in models with large $\tgb$ and small R-slepton masses, say $m_R$ around $100-200$ GeV. 
The planned improvements for $d_e$ would allow to test also models with $m_R$ up to the TeV region 
and moderate $\tgb$. 
When present or planned experimental limits turns out to overlap with these allowed ranges, 
bounds on the imaginary part of the relevant combinations of neutrino Yukawa couplings are obtained 
and plotted in the plane $(\ti M_1, m_R)$, respectively the bino and the average R-slepton masses at low energy. 
These plots allow to check whether any particular seesaw model is consistent with present data and, 
if so, which level of experimental sensitivity would test it.
In any case, an experimentally interesting contribution requires hierarchical 
right-handed neutrino masses.  

This feature is no more necessary when, in addition to the seesaw, a stage of $SU(5)$ grand unification is 
present above the gauge coupling unification scale. It is well known that the main drawback of minimal 
$SU(5)$ is that the value of the triplet mass, $M_T$, required by gauge coupling unification
is sizeably below the lower bound on $M_T$ derived from proton lifetime 
(see for instance \cite{isa_ijmpa} for recent reviews and references).
In our analysis we therefore keep the triplet mass as a free parameter. We nevertheless exploit the
minimal $SU(5)$ relations between the doublet and triplet Yukawa couplings since in general they are mildly
broken in non-minimal versions of $SU(5)$.
The simultaneous presence of right-handed neutrinos and heavy triplets turns out to further enhance 
the amplitude with the LL RR double insertion over the one with ${\cal I}m(A_{ii})$, essentially unaffected by
triplets. 
This cannot be derived by the - otherwise elegant - technique based on the 
allowed invariants \cite{romstr}. 

When right-handed neutrinos and Higgs triplets are simultaneously present, 
the predicted range for the radiatively-induced $d_e$ is already sizeably excluded by the present 
experimental limit. This in turn is translated into a strong constraint on the imaginary part
of the combination of Yukawa couplings which is relevant for the LL RR amplitude.
Such a constraint could be hardly evaded even for large values of the triplet mass and  
unfavorable supersymmetric mass parameters. 
The radiatively-induced $d_\mu$ in the presence of triplets should not exceed $10^{-23}$ e cm,
as was the case for the pure seesaw. However, at difference of the latter case, in the former one
$d_\mu$ only mildly depends on the spectrum of right-handed neutrinos.  
Notice that planned searches for $d_\mu$ could constrain (depending on the triplet mass mass, of course) 
the imaginary part of the combination of Yukawa couplings whose absolute value
could be independently constrained by the LFV decay $\tmg$.

The paper is organized as follows.
In Section 2 we introduce our notations, discuss the framework and make some preliminary considerations.
Section 3 considers the pure seesaw case by separately analyzing the flavour conserving (FC) and 
flavour violating (FV) amplitudes contributing to lepton EDMs. 
In Section 4 the framework of seesaw and $SU(5)$ is discussed along the same lines.
Concluding remarks are drawn in Section 5. Finally, in Appendix A and B we collect the RGE 
in the case of the seesaw, without and with a minimal $SU(5)$ unification respectively.


\section{Framework and Method}  \label{sec2}

In this section we recall the expression for the lepton EDMs in the mass insertion approximation
\cite{massins, moroi, prs, fms, ms1} 
to display the supersymmetric mass parameters that are constrained by 
the present and projected searches for $d_e$ and $d_\mu$.  
We then draw some preliminary considerations to introduce our procedure to calculate the radiative
1-loop contribution to these mass parameters. 
The relevant RGE can be found in the Appendix.

We adopt here the following conventions for the $6 \times 6$ slepton mass matrix in the 
lepton flavour (LF) basis where the charged lepton mass matrix, $m_{\ell}$, is diagonal: 
\beq
\left( \begin{array}{cc} {\tilde \ell}_L^\dagger & 
{\tilde \ell}_R^\dagger \end{array} \right) \ \ 
\left( \begin{array}{cc} m_{LL}^2 & A_e^\dagger v_d - \mu \tan\beta m_{\ell}   \cr
A_e v_d - \mu^* \tan\beta m_{\ell}  & m_{RR}^2  \end{array} \right)
\ \ \left( \begin{array}{c} {\tilde \ell}_L  \cr {\tilde \ell}_R 
\end{array} \right) \label{sleptonm2}
\eeq 
where $A_e$ is the $3 \times 3$ matrix of the trilinear coupling, the $A-$term.
All deviations from alignment in this mass matrix are gathered in the $\d$ matrices, 
which contain 30 real parameters (including 12 phases) and are defined as:
\bea
m^2_{LL}=m_L^2 (\mathbb{I} + \delta ^{LL} )  \quad & &\quad m^2_{RR}=m_R^2 ( \mathbb{I}+ \delta ^{RR} ) \nn\\
A_e^\dagger v_d - \mu \tan\beta m_{\ell}  &=& (A_{\ell \ell}^*  v_d - \mu \tan\beta m_{\ell} ) + m_L m_R \delta ^{LR} 
\eea
\noindent where $m_L , m_R$ are average masses for L and R sleptons respectively and 
$A_{\ell \ell}  \sim O(m_{susy} m_\ell/ v_d )$ are the diagonal elements of $A_e$, so that  
$\delta ^{LR}$ has only non-diagonal, flavour violating, elements.

The supersymmetric contributions to $d_i$, $(i=e,\mu,\tau)$, can be splitted in two parts,
involving respectively only flavour conserving (FC) or flavour violating (FV) elements of the slepton mass
matrix (\ref{sleptonm2}):  
\beq
d_i = d_i^{FC} +d_i^{FV} 
\eeq
\bea
d_i^{FC}  &= & \frac{e}{2}  \frac{\alpha M_1 }{4 \pi |\mu |^2 \cos^2 \theta _W} 
        [   m_i {\cal I}m(\mu) \tgb ( I_B + \frac{1}{2} I_L - I_R + I_2 ) - v_d {\cal I}m(A_{ii}^*) I_B ]  \label{FC}\\
d_i^{FV}  &= & \frac{e}{2}  \frac{\alpha M_1 }{4 \pi |\mu |^2 \cos^2 \theta _W} 
       \left[    m_R m_L (  {\cal I}m(\d^{LL}  \d^{LR})_{ii}  I'_{B,L} + {\cal I}m(\d^{LR}  \d^{RR} )_{ii}  I'_{B,R}  )  \right. \nn\\  
        && ~~~~~~~~~~~~\left.     +   \tgb (  {\cal I}m(\d^{LL} \mu \eta_\ell  m_\ell \d^{RR})_{ii}  I''_B  +
                                             {\cal I}m(\d^{LR} \mu^* \eta_\ell^*  m_\ell  \d^{LR})_{ii}  I''_B)  \right]            
\label{FV} 
\eea
\noindent where $\eta_{\ell} \equiv  \i3 -  A_{\ell \ell} v_d/ (m_{\ell}\mu \tan \beta)  \label{Mapp}$, 
the functions $I$ are defined as in \cite{ms1} and 
terms that are less relevant \footnote{E.g. a contribution to $d_i^{FC}$ of the form of (\ref{FV}), 
with $\d^{LR} \rightarrow $ $(A_{\ell \ell}^* v_d - \mu \tan\beta m_{\ell})/(m_L m_R)$,
is in principle present, but it is negligible with respect to (\ref{FC}).} 
or higher order in the $\d$'s matrix elements are omitted. Notice that $\eta_\ell \approx \i3$ for 
relatively large values of $\tgb$, favored in mSUGRA and for which the LF and CP violations are most 
likely to be detected. 
The FC and FV contributions could result from different seeds of CP violation but could also 
be correlated in many different ways in models. 
Anyway, the experimental limit can be put on both because, due to the different nature of 
the many parameters involved, an eventual cancellation between these contributions appears 
unnatural.
The distinction between the FC and FV contributions is also phenomenologically relevant: 
some of the $|\delta |$'s in the FV terms are already constrained to be smaller than $O(1)$ by LFV decays,
while ${\cal I}m(\mu)$ and ${\cal I}m(A_{i i})$ are directly constrained by the EDMs. 

Our aim here is to estimate the radiative contribution to the lepton EDMs induced by the seesaw interactions, 
first alone and subsequently accompanied by a stage of $SU(5)$ GUT.
Since the present experimental bounds on LFV decays and EDMs already point 
towards family blind soft terms (\ie~sparticles with the same quantum numbers must have the same soft terms) 
with very small CP phases, at the cut-off $\Lambda = M_{Pl}$ corresponding to the decoupling of 
gravitational interactions we assume real and flavour blind soft terms, namely in eq. (\ref{sleptonm2}),
\beq
m^2_{LL} = m^2_{RR} = \m02 \i3 ~~~, ~~~~\quad  A_e = y_e a_0~~~~~, 
\eeq
with real $a_0$, $m_0$ and $\mu$ term.
In the next sections we separately study the FC and FV contributions to $d_i$ and  
obtain explicit approximate expressions for ${\cal I}m(A_{i i})$ and the products of $\d$'s in (\ref{FV}). 
The effects of a more general family independence assumption are important 
but not crucial and can be easily included in our analysis.
By means of these general approximated expressions we will: \\
\noindent i) derive the upper prediction for the radiatively induced leptonic EDMs, stressing the model 
dependences;\\
\noindent ii) compare it with the experimental limits; \\
\noindent iii) when allowed by the experiment, obtain an upper bound on the imaginary part of the relevant 
combination of Yukawa couplings.

A couple of preliminary considerations are in order before presenting the results of the next
sections.

\subsection{On the naive scaling relation}

In the limit that all slepton masses are family independent, the FV contribution vanishes and the  
FC one is proportional to the mass of the $i-$th lepton (${\cal I}m (A_{ii}) v_d \approx {\cal I}m (a_0) m_{i}$)
leading, except an accidental cancellation 
\footnote{For dedicated studies on cancellations between amplitudes, also in more general
frameworks, see e.g. \cite{prs, cancell}.}
with the $\mu$-term amplitude, to the "naive" scaling relation 
\beq d_i/d_j = m_i/m_j~~~~. \label{naive} \eeq 
Then, due to the present experimental limit on $d_e$, 
$d_\mu$ could not exceed $d_e m_\mu/m_e  \sim 2 \cdot 10^{-25}$ e cm, which roughly corresponds to the 
planned sensitivity and, 
if the limit on $d_e$ were still to be lowered, next generation experiments would have no 
chance of measuring $d_\mu$.

Such considerations could provide interesting informations because (\ref{naive})
strictly apply only to the FC $\mu$-contribution to the EDM, while in general  
both ${\cal I}m (A_{ii})$ and the FV terms may strongly violate it. 
This is the case for the radiatively induced ${\cal I}m (A_{ii})$. 
Some of the FV contributions are instead naturally proportional to a different lepton mass, $m_k$,  
possibly heavier than $m_i$, as discussed in \cite{fms, romstr} - and, before, for the quark sector, in \cite{bs}. 
In particular, it will turn out in the next sections that the FV contribution can even take over the FC one. 
Hence, a value of $d_\mu$ above $\sim 2 \cdot 10^{-25}$ e cm is a possibility that deserves 
experimental tests and, interestingly enough, it would imply the source of lepton EDM being either the FV 
contribution or a non-universal ${\cal I}m(A_{ii})$, 
so providing a remarkable hint for our understanding of CP violation.

\subsection{On the relevant combination of Yukawa couplings}

Once a theory is specified, it is relatively easy to list the combinations of Yukawa couplings 
appearing in the slepton mass radiative corrections that may contribute to lepton EDMs. 
However, the actual calculation is more involved as we now turn to discuss.

As an example, let us look for ${\cal I}m(A_{ii})$ by studying 
the evolution of $A_e$ in the case of degenerate right-handed neutrino masses, 
$\bar M$, for simplicity. 
Let us first consider the case of the pure seesaw and 
define \footnote{Here and in the following Dirac mass terms are always written as $\bar f_R m_f f_L$.} 
the hermitian matrices $E \equiv y_e^{\dagger} y_e$, $N \equiv y_{\nu} ^{\dagger} y_{\nu}$.
Notice that $E$ is real and diagonal in the LF defining basis and $N$ is diagonalized
by a unitary matrix similar to the CKM one with only one phase (even for non-degenerate right-handed neutrinos).
By solving the RGE for $A_e$, eq. (\ref{Ae_ss}), linearly in $t_3 \equiv 1/(4 \pi)^2$ $\ln(M_{Pl}/\bar M)$, 
$A_e$ can only be proportional to $y_e E$ and $y_e N$, whose diagonal elements are real. 
At $O(t_3^2)$ only $y_e EE, y_e NN, y_e NE$ and $y_e EN$ appear, 
whose diagonal elements are again real. 
A potential ${\cal I}m(A_{ii})$ shows up at $O(t_3^4)$, through ${\cal I}m(y_e N [N,E] N)_{ii}$.
On the other hand, when also Yukawa interaction of the $SU(5)$ triplets are present, eq. (\ref{Ae_su5})
shows that, at first order in $t_T \equiv 1/(4 \pi)^2$ $\ln(M_{Pl}/ M_T)$, 
$A_e$ can be proportional to $y_e E, y_e N$ but also to $U^* y_e$, where $U \equiv y_u^\dagger y_u$. 
The latter have real diagonal elements but allow at $O(t_T^2)$ for a combination
with diagonal imaginary part, namely $(U^* y_e N)_{ii}$.

Accordingly, in order to evaluate the actual coefficients in front of the products of Yukawa 
coupling matrices which are likely to have phases,
we solve the RGE for the soft parameters by a Taylor expansion in the
small parameters $t_{if}=1/(4 \pi)^2 \ln(Q_i/Q_f)$ associated to the intervals between the 
successive decoupling thresholds of the various heavy states.
For instance, by integrating the RGE for $A_e$ in the case of $SU(5)$ plus seesaw, 
a non vanishing coefficient is obtained for the combination ${\cal I}m(U^* y_e N )_{ii}$ at $O(t_T^2)$.
Now, lepton FV transitions and CP phases are naturally defined in the LF basis where $y_e$ and the Majorana 
masses $M_R$ are diagonal and real. This basis is not invariant under the RGE evolution and 
one should diagonalize $y_e$ and $M_R$ again at the lower scale. 
Therefore, one has to find out the effect of these final rotations.

We adopt the rotating basis method introduced 
in Ref. \cite{bs}, where the RGE are modified to incorporate the fact that the matrices are defined in the LF
basis at each scale, so that $y_e$ and $M_R$ are always diagonal. 
It is worth to stress that, even if the rotations needed to diagonalize $y_e$ are small, 
their effect could be crucial for ${\cal I}m(A_{ii})$ and the four products of $\d$'s in eq. (\ref{FV}).
For instance, in the case of $SU(5)$ plus seesaw it turns out that this correction exactly cancels the term 
proportional to $t_T^2  {\cal I}m( U^* y_e N )_{ii}$ in $A_e$. 
On the contrary, these rotations can be safely neglected when deriving approximate expressions 
for the radiatively-induced LFV decays because the latter are only sensitive to absolute
values of $\d$'s.


\section{Lepton EDMs and Seesaw}

In this section we consider the predictions for lepton EDMs in the context of the 
supersymmetric extension of the seesaw model, namely we assume the MSSM supplemented with the 
seesaw Yukawa interactions and Majorana masses for the right-handed neutrinos as the effective theory 
valid up to the cut-off $\Lambda = M_{Pl}$ where gravitational interactions decouple. 
Starting with real and universal boundary conditions at $M_{Pl}$, 
we solve the RGE displayed in Appendix A
by expanding in the small parameters defined by the right-handed neutrino thresholds 
\beq
t_3= \frac{1}{(4 \pi)^2} \ln \frac{\Lambda}{M_3} ~~~~~\quad 
t_2= \frac{1}{(4 \pi)^2} \ln \frac{M_3}{M_2} ~~~~~~\quad 
t_1= \frac{1}{(4 \pi)^2} \ln \frac{M_2}{M_1}  ~~~ ,\label{thresholds}
\eeq
with the ordering $M_3 > M_2 >M_1$.
We thus obtain approximate analytic expressions for the radiatively induced $\delta$'s and ${\cal I}m(A_{ii})$  
which depend on $m_0$, $a_0$ and the Yukawa couplings defined in the LF basis at the scale $\Lambda$.  
Of course the latter can be immediately translated into the corresponding ones defined 
at any scale, \eg  $M_1$ or $m_{susy}$. 

Precisely because of the lepton flavour and CP violating Yukawa couplings, 
the LF basis is continuously rotated and rephased with the RGE evolution and,   
as already discussed, we handle this by working in a rotating basis. It turns out that
the basis transformation introduces negligible corrections in the FC terms (\ref{FC})
but important ones in the FV terms (\ref{FV}). 
Notice also that the seesaw effects stop at the decoupling threshold of the lightest right-handed neutrino, $M_1$,
and that the RGE evolution of these effects down to the supersymmetric scales where CP and lepton FV transitions 
are estimated is generically small and can be neglect for our estimates. 
We now study separately the FC and the FV contributions.


\subsection{Flavour conserving contribution}

Starting in the LF basis at $M_{Pl}$ from $A_e =  y_{e} a_0$,
the seesaw interactions generate a ${\cal I}m(A_{ii})$ in the LF basis at $M_1$. 
At leading order in the $t$'s 
and defining $y_\nu^\dagger P_a y_\nu \equiv \a$  ($a=1,2,3$) and the
right-handed neutrinos projectors $P_1={\rm diag}(1,0,0)$, $P_2={\rm diag}(1,1,0) $,  $P_3 = \i3$,
the latter reads \footnote{Actually, (\ref{ss_ae}) is an approximation by excess and improves as the involved
Yukawa couplings become small. However, for our estimates it is reliable up to $O(1)$ Yukawa couplings.}:
\beq
{\cal I}m (A_{ii}) =   8 a_0 y_{e_i}  \left( t_2 t_3  {\cal I}m( \2  \3)_{ii} + t_1 t_3 {\cal I}m(\1  \3 )_{ii} + 
 t_1 t_2  {\cal I}m( \1 \2 )_{ii}   \right) ~~, 
\label{ss_ae}
\eeq
where the Yukawas in the r.h.s. are evaluated at $\Lambda$.
A similar  
\footnote{As far as the comparison with the expressions in Refs. \cite{ellis-1, ellis-2, ellis-n} is possible, 
we find agreement with the last papers up to the coefficient of ${\cal I}m( \1 \2 )_{ii}$.}
formula were previously presented in Refs. \cite{ellis-1, ellis-2, ellis-n}, 
but with the various Yukawas involved defined at different scales. 
Then, one must be more cautious in drawing general conclusions and the authors validate theirs 
with specific numerical examples. 
Our approach offers the advantage that the corresponding results become more transparent 
and allow for an easier estimate of the effects once a pattern of seesaw parameters is assigned.
Anyway, the crucial point \cite{ellis-1} is that the more right-handed neutrinos are hierarchical, 
the more the FC contribution increases. 
Indeed, it vanishes in the limit that right-handed neutrinos are degenerate, in which case a contribution to
${\cal I}m(A_{ii})$ only appears at fourth order, proportional to ${\cal I}m(y_e N[N,E]N)_{ii}$. 
Notice also that the naive scaling relation is generally violated \cite{ellis-1} by (\ref{ss_ae}).

\begin{figure}[!h]
\vskip .5 cm
\centerline{
\psfig{file=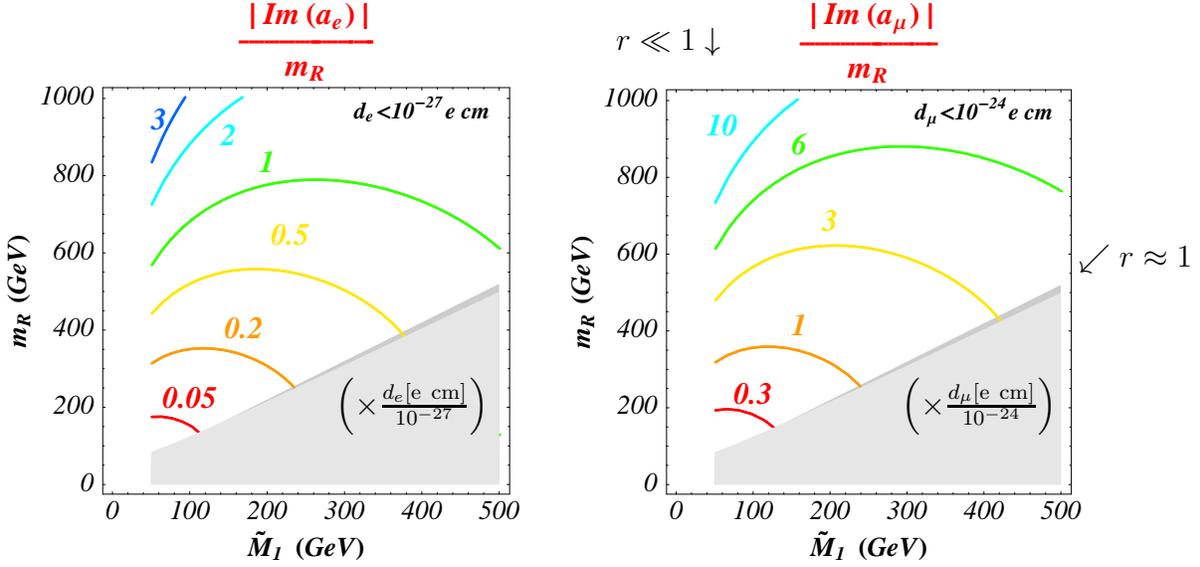,width=1 \textwidth}
\put(-250,250){\Large Experiment}
\put(-190,200){$ r \ll 1 \downarrow $}
\put(-15,118){$ \swarrow r \approx 1  $}
\put(-300,62){ $ \left( \times \frac{d_e [{\rm e~cm}]}{10^{-27}} \right) $} 
\put(-86,62){ $ \left( \times \frac{d_\mu [{\rm e~cm}]}{10^{-24}} \right) $ } }
\caption{Experimental upper bound on $| {\cal I}m (a_e) |/m_R$ and $| {\cal I}m (a_\mu) |/m_R$ for
$d_e < 10^{-27}$ e cm and $d_\mu < 10^{-24}$ e cm respectively \cite{ms1}. $r^{-1}$ corresponds to the
slope: $r \equiv \ti M_1/m_R$. For the present sensitivity $d_\mu < 10^{-18}$ e cm, the numbers have to
be multiplied by $10^{6}$. $m_L$ is fixed as in mSUGRA.}
\label{Fph_A}
\end{figure}

Eq. (\ref{ss_ae}) displays a linear dependence on the unknown parameter $a_0$. 
By defining $A_{ii} \equiv |A_{ii}| e^{i 8 \phi_{A_i}}$, then 
$ \phi_{A_i} \approx  t_2 t_3  {\cal I}m( \2  \3)_{ii} + $ $t_1 t_3 {\cal I}m(\1  \3 )_{ii} + $ 
$ t_1 t_2  {\cal I}m( \1 \2 )_{ii} $ is completely specified by the seesaw parameters. 
However, as appears from eq. (\ref{FC}) the experimental limit on $d_i$ doesn't probe directly $\phi_{A_i}$,   
rather it gives a bound on ${\cal I}m(A_{ii}) v_d \equiv m_{i} {\cal I}m(a_{i})$ 
once supersymmetric masses are fixed (and up to unnatural conspiracies between the various amplitudes). 
Figs. \ref{Fph_A}, taken from Ref. \cite{ms1}, show
the present upper limits on $| {\cal I}m (a_e) |/m_R$ and the planned ones for $| {\cal I}m (a_\mu) |/m_R$ 
in the plane $(\ti M_1, m_R)$, respectively the bino and R slepton mass at $m_{susy}$.
Notice that these limits are quite model independent because, apart from $m_R$ and $\ti M_1$, 
there is only a mild dependence on $m_L$, which we have fixed as in mSUGRA for definiteness.
Indeed, since the $A$-term amplitude in $d_i^{FC}$ arises from pure bino exchange, it does not involve 
\footnote{The function $I_B$ in eq. (\ref{FC}) contains a factor $|\mu|^2$ that cancels
the one in the overall coefficient. The exact expressions for $I_B$ can be found in \cite{ms1}, as well as
approximate ones suitable for various pattern of supersymmetric masses.} 
$\mu$ nor $\tan \beta$. 
In mSUGRA 
there is an unphysical region in the plane $(\ti M_1, m_R)$ corresponding to $m_0^2 < 0$ and which has been 
indicated in light grey in the plots. Anyway,  
the dark grey region and below is also excluded because $m_R  \le \ti M_1$, 
in contrast with the requirement of neutrality for the LSP.

To find an upper estimate for the seesaw induced $d_i^{FC}$, let us evaluate
$| {\cal I}m (a_i) |/m_R$ from eq. (\ref{ss_ae}) by considering only its first term. 
This situation is representative because the terms proportional to $t_1$ are negligible when the 
lightest right-handed neutrino has smaller Yukawa couplings, as happens in many models and 
as one would guess from similarity with the charged fermion sectors. 
Anyway, the following discussion is trivially adapted to other cases.
For definiteness, we adopt this set of reference threshold values:
$ M_2=10^{12}$ GeV, $M_3=10^{15}$ GeV, $\Lambda=2~10^{18}$ GeV. 
Since, as demanded by perturbativity, $(\2  \3)_{ii} \le O(1)$,
\beq
\frac{    {\cal I}m (a_i) }{m_R}  \le O \left( 0.02  \frac{t_2 t_3}{2~ 10^{-3}}    \right)  \frac{a_0}{m_R}     ~~~ .
\label{ae_ss}
\eeq
This upper estimate is easily adapted to any given model once the relation between 
$a_0$ and $m_R$ is made explicit. 
To make the dependence more manifest, let us introduce the following two mSUGRA
situations: a)  $a_0^2 = 2 m_0^2$; b) $a_0^2 =\ti  M_{1/2}^2$, where $\ti M_{1/2}$ is the universal gaugino
mass at the gauge coupling unification scale. 
In the first case $(a_0/m_R)^2 \approx 2 (1-0.9 r^2)$, with $r \equiv \ti M_1/m_R$.
The ratio $a_0/m_R$ thus displays a mild excursion as it decreases from $\sqrt{2}$ down to $0.45$ 
when moving from the vertical left axes, where $r \ll 1$, to the joining line of the dark grey region, where $r =1$. 
For case b), the situation is opposite and not so mild: $(a_0/m_R) \approx 2.5~ r$. 
In the more natural cases in between, the ratio $a_0/m_R$ should thus be rather stable.

Let us now compare the upper estimate (\ref{ae_ss}) with the experimental bound.
As appears from eq. (\ref{FC}), the latter improves linearly with the 
experimental sensitivity to $d_i$. Taking $a_0 \sim m_R$, figs. (\ref{Fph_A}) show that values around $0.02$ for 
$| {\cal I}m (a_{i}) |/m_R$ would require an experimental sensitivity to $d_e$ and $d_\mu$ respectively
at the level of $10^{-28}-10^{-29}$ e cm and $2~10^{-26}-2~10^{-27}$ e cm, the exact value depending on the
particular point of the $(\ti M_1, m_R)$ plane. 
However, to avoid charge and color breaking, the constraint $a_0/m_R \le 3$ in general applies. 
Thus, focusing for instance around $m_R \approx 500$ GeV, the FC contribution  
cannot exceed $\sim10^{-28}$ e cm for $d_e$ and $\sim 2~ 10^{-26}$ e cm for $d_\mu$,
even with highly hierarchical right-handed neutrinos.
Notice that the former value could be at hand of future experimental searches for $d_e$, 
while the latter is at the very limit of the planned experimental sensitivity to $d_\mu$.
Allowing for smaller $m_R$ values, $m_R \approx 200$ GeV, the FC seesaw upper estimate 
comes close to the present limit for $d_e$ while it cannot be more than
$\sim 2~10^{-25}$ e cm for $d_\mu$. 

The result for $d_\mu$ is thus a kind of "negative" one, since its eventual future discovery above
$\sim 2~10^{-25}$ e cm couldn't be attributed to the radiative FC contribution of the seesaw.
Even if the situation for $d_e^{FC}$ appears more optimistic, it is worth to underline that the 
previous upper estimate actually applies to models with at least four $O(1)$ neutrino Yukawa 
couplings and large CP phases, like those in Ref. \cite{ellis-n}. 
However, as we now turn to discuss, the FV contribution, underestimated by previous 
analyses, could drastically enhance the $d_i$ upper estimate.


\subsection{Flavour violating contributions}

In the following we isolate the potentially most important products of $\d$'s and give an estimate of their
relative magnitude with respect to the FC amplitude.
The relevant approximations for the flavour violating elements, $i \neq j$, of the $\d$'s are: 
\bea
m_L m_R \d^{LR}_{ij} & = & a_0 m_{i} 
[  - 2  t_3 {\cal N}_{ij}    +  \sum_{a} \frac{t_a^2}{2} F_A(a , a )_{ij} +  \sum_{a>b} t_a t_b F_A(a,b)_{ij} ]
\label{eq_Aess} \\
m^2_R \d^{RR}_{ij}  & = & \sum_{a}  \frac{t_a^2}{2} F_R( a , a )_{ij}  +  \sum_{a>b}   t_a t_b F_R(a,b)_{ij} 
\label{eq_Rss} \\
m^2_L \d^{LL}_{ij}  &=&  - (6 \m02 +2 \A02 ) t_3 {\cal N}_{ij}  
                                  +  \sum_{a}  \frac{t_a^2}{2} F_L( a , a )_{ij} +  \sum_{a>b}  t_a t_b F_L(a,b)_{ij}    
\label{eq_Lss}
\eea
where $a,b=1,2,3$ , the matrix $\cal N$ is defined as
$ t_3 {\cal N}  \equiv t_3 \3 + t_2 \2 +t_1 \1 $ and
\bea
F_A(a,b) &=&  15   \{ E , \a \} - 5  \{ E,  \b  \}  + 12  \{ \a,  \b  \} + 4 [\b, u_E^{(b)} ] \nn\\
       &+& 2 ( ( {\b}_{(d)} + D_e ) \a +( {\a}_{(d)} + D_e ) \b )  +4(\ti D_\nu^{(a)}/a_0 +D_\nu^{(a)}) \b \nn\\
&+&  [\a,E] + 4 [\b,\a] + 7 [E,\b]   ~~, \\
F_R(a,b) & = &  8 [  (6 \m02 + 4 \A02) y_e \a y_e^\dagger -  (6 \m02 +2 \A02 ) y_e \b y_e^\dagger  ] 
\label{FR_ss} \\
F_L(a,b) & =& 2 (6 \m02 +2 \A02 ) ( \{ 3 \b + E, \a  \} +  2 D_\nu^{(a)} \b  + [\a,u_E^{(b)}] ) +2   {m^2}'_{H_{u}} \b \nn\\
               & +& 4 \A02 ( \{ 3 \b +E , \a \}+\{ E,\b \})  + 2 (G_L+4 a_0 \ti D_\nu^{(a)}) \b 
\label{FL_ss}
\eea       
with the Yukawas in the r.h.s. evaluated at the scale $\Lambda$. 
The subscript $(d)$ indicates to take only the diagonal elements of the matrix, 
$u_E^{(a)}$ is defined through $[u_E^{(a)},E]=\{E, 3 E+ \a + D_e \}$ and the definition
of all the other quantities can be found in Appendix A.  

By means of the above expressions, the predictions for the imaginary part of the various 
product of $\d$'s can be studied.
An imaginary part in the products $(\d^{LL} \d^{LR})_{ii}$, $(\d^{LR} \d^{RR})_{ii}$, ...,
only arises at third order in at least two different $t$'s.

\subsubsection{The LL RR contribution}

Let us firstly discuss the LL RR double insertion, ${\cal I}m(\d^{LL} \mu \eta_\ell m_\ell \d^{RR})_{ii}$, 
which turns out to be the quantitatively most interesting one.
Since in general $\eta_\ell \approx \i3$ and the phase of $\mu$ - if any -  is extremely small so that it can 
be safely neglected in the discussion,
${\cal I}m(\d^{LL}  \mu \eta_\ell m_\ell \d^{RR})_{ii}   \approx   \mu  {\cal I}m(\d^{LL}m_\ell  \d^{RR})_{ii}$.
From eqs. (\ref{eq_Rss}, \ref{eq_Lss}) it appears that $\d^{RR}_{ij}$, $\d^{LL}_{ij}$ are respectively 
of second and first order in the $t$'s. Thus, the
lowest order is the cubic and the LL RR contribution reads:
\beq
{\cal I}m(\d^{LL} m_\ell  \d^{RR})_{ii}  = 
8  m_{i}   \frac{ (6 \m02 + 2 \A02 )  (6 \m02 + 3 \A02 ) }{m^2_L m^2_R}  
\sum_{a>b} t_a t_b (t_a+t_b)    {\cal I}m(N_a E N_b)_{ii}~.
\label{ss_LLRR}
\eeq
Notice that ${\cal I}m(N_a E N_b)_{ii} = \tan^2 \beta {\cal I}m(N_a m_\ell^2 N_b)_{ii} /m_t^2$, where $m_t$ is the top
mass. Due to the hierarchy in $m_\ell$, a potentially important effect could only come from 
${\cal I}m({N_a}_{i3}  {N_b}_{3i})$.
Notice also that, on the contrary of the FC contribution, (\ref{ss_LLRR}) doesn't strongly depend on 
$a_0 $. 

\begin{figure}[!h]
\centerline{
\psfig{file=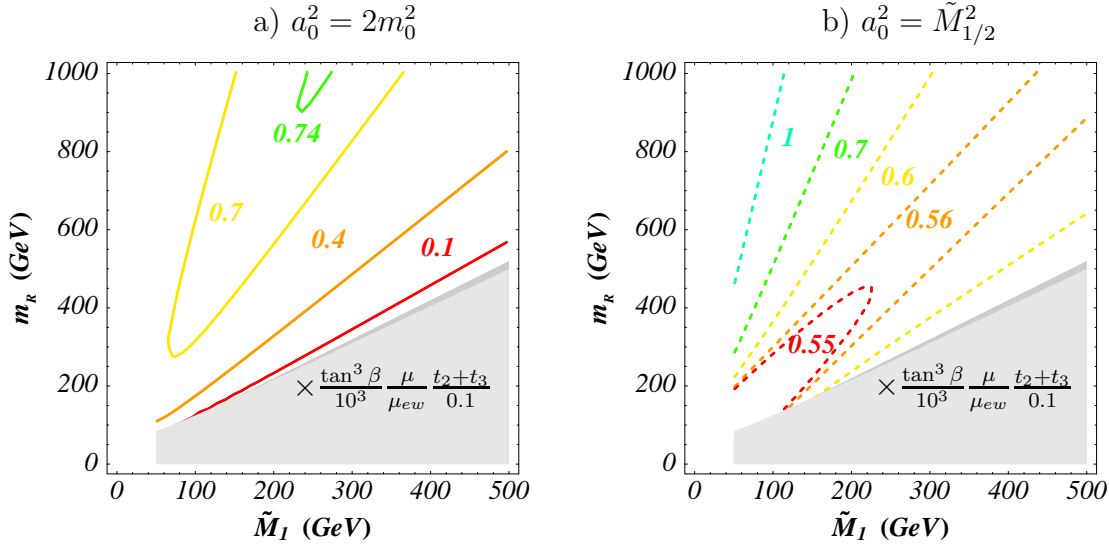,width=1 \textwidth}
\put(-300,235){ \Large Theory:  $\frac{ d_i^{FV_{LLRR}} } { d_i^{FC} }$ for Seesaw} 
\put(-335,200){ a) $a_0^2=2 m_0^2$}   \put(-120,200){ b) $a_0^2= \ti M_{1/2}^2$}
\put(-320,62){ $ \times \frac{\tan^3 \beta}{10^3} \frac{\mu}{\mu_{ew}} \frac{t_2+t_3}{0.1}$} 
\put(-100,62){ $\times \frac{\tan^3 \beta}{10^3} \frac{\mu}{\mu_{ew}} \frac{t_2+t_3}{0.1}$ } }
\caption{Ratio $d_i^{FV_{LLRR}}/d_i^{FC}$ when 
${\cal I}m(\3 m_\ell^2 \2)_{ii} \approx m_\tau^2 {\cal I}m(\3 \2)_{ii}$. We have assumed as reference: 
$\Lambda/M_2 = 2~10^{6}$, $m_L$ as in mSUGRA, $\mu=\mu_{ew}$ and $\tgb=10$. }
\label{Fss_R}
\vskip 1 cm
\end{figure}

The relative importance between this amplitude and the FC one can be easily appreciated 
by neglecting the terms proportional to $t_1$, which, as already mentioned, 
is justified when the lightest right-handed neutrino has the smallest Yukawa couplings:
\beq
\frac{d_i^{FV_{LLRR}}}{d_i^{FC}} = \frac{\mu \tan^3 \beta}{a_0} \frac{ (6 \m02 + 2 \A02 ) 
(6 \m02 + 3 \A02) }{ m_R^2 m^2_L}
(t_2 +t_3) \frac{{\cal I}m(\3 m_\ell^2 \2)_{ii}}{ m_t^2 ~{\cal I}m(\3  \2)_{ii}}   \frac{I''_B}{I_B}~.
\eeq
For realistic values, the ratio of the two loop functions is slightly smaller than one.
To obtain a rule of thumb, let us take $m_0  \sim a_0 \sim m_{L,R}$ 
and the reference ratio $\Lambda/M_2 = 2~10^{6}$. 
Then, unless ad hoc fine-tunings in the structure of $N$,
${\cal I}m(\3 m_\ell^2 \2)_{ii} \sim m_\tau^2 {\cal I}m(\3 \2)_{ii}$, and one finds 
\beq
\frac{d_i^{FV_{LLRR}}}{d_i^{FC}} \sim 0.5 ~\frac{\tan^3 \beta}{10^3} \frac{\mu}{\mu_{ew}} \frac{t_2+t_3}{0.1}
~~~~~~~~~~~{\rm (rule ~of ~thumb),}
\label{rot} \eeq
where $|\mu_{ew}|^2 \approx 0.5 m_R^2 + 20 \ti M_1^2$ is the value of $\mu$ accounting for radiative
electroweak breaking in mSUGRA.
Despite being of third order in the $t$'s,
the FV amplitude can take over the FC one thanks to its $\tan^3 \beta$ dependence.
The precise value of this ratio is displayed in figs. \ref{Fss_R} in the plane $(\ti M_1, m_R)$.
Cases a) and b) are separately displayed 
so that the behavior for any situation in between can be easily extrapolated. 
The relevant generalizations are also reminded.
The plots show that the rule of thumb is quite reliable and allow to extract, for each point of the plane, 
the value of $\tgb$ for which the FV amplitude takes over the FC one.
In both cases a) and b), this happens for $\tan \beta \gtrsim 10$ - the only exception
being the region with $r \approx 1$ for case a), where $\tgb \gtrsim 20$ is required.

\subsubsection{The other contributions}

The amplitudes with the other products of $\d$'s are less important than the LL RR one. 
Consider for instance ${\cal I}m(\d^{LR} \d^{RR})_{ii}$. 
The flavour violating elements in $\d^{RR}$ and $\d^{LR}$ are respectively of second and first order,
so that the product appears at third order:
\beq
{\cal I}m ( \d^{LR} \d^{RR} )_{ii} = 8 \frac{ m_{i} a_0   (12 \m02 + 6 \A02) }{m_R^3 m_L} \sum_{a>b}  
t_a t_b (t_a+t_b) {\cal I}m(N_a E N_b)_{ii}~. 
\eeq
It vanishes in the limit $a_0 \rightarrow 0$, as the FC contribution.
Neglecting the terms proportional to $t_1$, the ratio of the LR RR amplitude and 
the FC one reads
\beq
\frac{d_i^{FV_{LRRR}}}{d_i^{FC}}= \tan^2 \beta \frac{12 \m02 + 6 \A02 }{m_R^2} (t_2 +t_3) 
\frac{{\cal I}m(\3 m^2_\ell \2)_{ii}}{m_t^2 {\cal I}m(\3  \2)_{ii}} 
\frac{I'_{B(R)}}{I_B}~.
\label{LRRR}
\eeq
It is easy to check that this ratio is smaller than one (unless ad hoc fine-tunings in the structure of $N$):  
the ratio of the two loop functions is slightly smaller than one for realistic values of the supersymmetric 
parameters and, taking for instance $m_0=a_0$ and $\Lambda/M_2 = 2 ~10^{6}$, 
(\ref{LRRR}) should not exceed $\sim~10^{-4} \tan^2 \beta$.
  
For ${\cal I}m(\d^{LL} \d^{LR})_{ii}$, no imaginary part can arise at second order in $t$'s because
both $ \d^{LL}$ and $\d^{LR}$ are proportional to $t_3 {\cal N}$.
At third order there are many contributions and it is lengthy but straightforward to check that they 
are proportional to at least two different $t$'s. This contribution is also proportional to $a_0$ and
could be comparable to the FC one but is in any case smaller than the LL RR one. 
The expression for the double LR insertion is also quite involved. 
It is proportional to $a_0^2$ and, being also suppressed by a factor $m_\ell^2 /m_L^2$ with respect to the 
LL RR one, it can be safely neglected.

\subsection{Predicted range and constrains on Yukawas}

For values of $\tan \beta \gtrsim 10$, for which the EDMs are enhanced and thus most likely
to be observed, the FV amplitude with the LL RR double insertion is generically dominant with respect
to all other amplitudes. 
Then, when $\tgb \gtrsim 10$, all the considerations made previously  
for $d_i^{FC}$ actually apply to $d_i$ when strengthened by the rule of thumb factor (\ref{rot}). 
To give an example, if $t_2+t_3 \sim 0.1$ and $\mu \sim \mu_{ew}$,
$d_e$ cannot exceed $\sim 0.5~10^{-28(-27)} ~\tan^3 \beta/10^3 $ e cm 
and $d_\mu$  $\sim 10^{-26(-25)}~\tan^3 \beta/10^3$ e cm when $m_R \sim 500 (200)$ GeV. 
Planned experimental sensitivities to $d_\mu$ could then test the seesaw radiative contribution
for models with small $m_R$ and/or large $\tan \beta$, in which case the factor (\ref{rot}) could 
be up to $\sim 50 - 60$ so that $d_\mu$ should not exceed $O(10^{-23})$ e cm. 
On the contrary, the range of the seesaw induced $d_e$ already overlaps
with the present experimental limit for values of $m_R$ up to $1/2$ TeV when $\tgb \sim 30$. 

\begin{figure}[!t]
\centerline{
\psfig{file=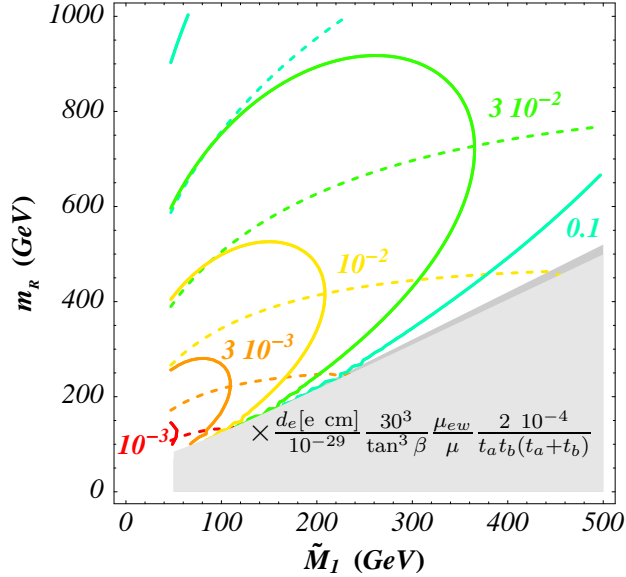,width=.6 \textwidth}
\put(-200,245){ {\small Upper bound on} ${\cal I}m({N_a} \frac{m_\ell^2}{m_\tau^2} {N_b})_{11}$  }
\put(-160,53){  $\times  \frac{d_e[ {\rm e ~cm} ]}{10^{-29}} \frac{30^3}{\tan^3 \beta}
 \frac{\mu_{ew}}{\mu} \frac{2 ~10^{-4}}{t_a t_b (t_a + t_b)} $ }     }
\caption{Upper bound on ${\cal I}m({N_a} m_\ell^2 {N_b})_{11}/{m_\tau^2}$. 
Solid (dashed) lines refer to case a) (b)).  } 
\label{Fss_INN} 
\end{figure}

Barring unnatural cancellations, planned searches for $d_e$ could thus test each term of the sum in
(\ref{ss_LLRR}), namely each \footnote{Needless to say, this remains true even if the FC
were the dominant amplitude, in which case the bound would be stronger.}
\beq
t_a t_b (t_a+t_b)    {\cal I}m(N_a \frac{m^2_\ell}{m^2_\tau} N_b)_{11}~~~~~~~~~~(a>b)~~~~.
\eeq
The effect of an eventual two orders of magnitude improvement for $d_e$
on the upper limit on ${\cal I}m(N_a m^2_\ell   N_b)_{11}/m^2_\tau$
is displayed in fig. \ref{Fss_INN} 
by taking as reference values $\tgb =30$ and $t_a t_b (t_a+t_b) = 2~10^{-4}$.
It turns out that planned limits could be severe enough to test models with hierarchical neutrino Yukawa couplings.
This cannot be done by present limits.
For any given seesaw model, it is straightforward to extrapolate from the plot 
the level of experimental sensitivity required to test it.  
Notice also that for $d_\mu$, 
limits on ${\cal I}m(N_a m^2_\ell   N_b)_{22}/m^2_\tau$
as strong as the present ones on 
${\cal I}m(N_a m^2_\ell   N_b)_{11}/m^2_\tau$ 
would require a sensitivity to $d_\mu$ at the level of $2~10^{-25}$ e cm.

\section{Lepton EDMs, Seesaw and $SU(5)$ Triplets }

We now add to the supersymmetric seesaw model a stage of a minimal $SU(5)$ GUT
above the gauge couplings unification scale, $M_{GUT} \sim 2~ 10^{16}$ GeV. 
Namely, we include the contribution of the higgs triplets Yukawa interactions to the RGE evolution of slepton
masses from $M_{Pl}$ down to their threshold decoupling scale $M_T$, which is very likely to 
be bigger than $M_3$.  
Notations are defined in Appendix B.

It is not restrictive to work (at any scale) in the basis where $(y_d^T=)y_e$ and $M_R$ are real and diagonal
and $y_u = V^T d_u \phi_u V$, where $d_u$ are the (real and positive)
eigenvalues of $y_u$, $V$ is the CKM matrix in the standard parameterization (more on this later)
and $\phi_u$ is a diagonal $SU(3)$ matrix.   
The RGE for the radiatively induced misalignments are written in eqs. (\ref{rsu5_y}) to (\ref{rsu5_A}). 
At first order in $t_{1,2,3}$ defined in eq. (\ref{thresholds}) and 
\beq
t_T \equiv \frac{1}{(4 \pi)^2} \ln\frac{M_{Pl}}{M_T}  ~~~,
\eeq
their solutions at the scale $M_1$ reads:
\bea
m^2_R \d^{RR}  &=& -   (6 \m02 +2  \A02 ) (2  t_{1}  y_e^2 + 3 t_T   U^*)  \nn\\
m^2_L \d^{LL} &=& -   (6 \m02 + 2 \A02 ) ( (3 t_T + t_{1} ) y_e^2 +  t_3 {\cal N} )   \label{eq_dsu5}\\ 
m_L m_R \d^{RL} &=& - a_0 (  6 t_T   U^*  m_\ell +  2 m_\ell  t_3 {\cal N}   )    \nn
\eea
where the matrix $\cal N$ is defined as $t_3 {\cal N}  \equiv t_3 \3 + t_2 \2 +t_1 \1 $.
It is understood that all the quantities in the r.h.s. of (\ref{eq_dsu5}) are evaluated at $\Lambda$.
The small effects of the subsequent evolution from $M_1$ down to $m_{susy}$ can be neglected in the following
discussion. 
We explicitly write only the first order terms
\footnote{Of course eqs. (\ref{eq_dsu5}) overestimate the misalignment. 
However, for our estimates here they are reliable up to  $y_t \sim y_{\nu_3} \sim 1$.}
in the $t$'s because, contrarily to the seesaw case, they already produce a potential imaginary part for the 
FV contribution, eq. (\ref{FV}).
Instead, for the diagonal part of $A_e$ this is not the case, as we now turn to discuss.


\subsection{Flavour conserving contribution}

As anticipated in the simplified discussion of Section \ref{sec2}, 
a potential candidate for ${\cal I}m(A_{ii})$, proportional to ${\cal I}m(U^* y_e N)_{ii}$, 
could show up at $O(t_T^2)$.
It is lengthy but straightforward to see that such term is exactly canceled by the effect of rotating the basis.
For the same reason, in the general case with different thresholds $M_{1,2,3}$, $M_T$, the overall coefficient
of ${\cal I}m(U^* y_e {\cal N})_{ii}$ is zero. 
Therefore, the second order contribution to ${\cal I}m(A_{ii})$ is just
\bea
{\cal I}m(A_{ii}) = 8 ~a_0~ y_{e_i}~  [ ~ 
                                     ( \frac{3}{2}~ t_T  +  ~t_3 ) t_2 ~ {\cal I}m(\2 \3)_{ii} 
                                  + ( \frac{3}{2}~ t_T  +  ~t_3 ) t_1 ~ {\cal I}m(\1 \3)_{ii}    \nn\\
                                   +   t_2 t_1  {\cal I}m(\1 \2)_{ii} ~]~~~ ~~~~,~~~~~~~~~~~~~~~~~~~~~~~~~~~~~~~~~~~~~~~~~\label{ae_ss+su5}
\eea
namely the pure seesaw one discussed in the previous section, eq. (\ref{ae_ss}),
with the substitution $t_3 \rightarrow (3/2~ t_T  +  ~t_3 )$, due to the fact that above $M_T$ also the 
triplets circulate in the loop renormalizing the wave functions. 
$t_T$ is naturally expected to be small (triplets will not decouple much below $10^{16}$ GeV) so that 
eventual higher order contributions involving $t_T$ are expected to be negligible with respect to (\ref{ae_ss+su5}).
As a result, a stage of $SU(5)$-like grand unification, cannot enhance by much the FC contribution 
with respect to the pure seesaw case.


\subsection{Flavour violating contribution}

On the contrary, products of two $\delta$'s have an imaginary part proportional to 
$t_3 t_T $ ${\cal I}m(U^* y_e {\cal N})_{ii}$ and 
the FV contribution to $d_i$ is potentially bigger than the FC one.
Most interestingly, the predicted range for the radiatively induced $d_e$ turns out to have been 
already sizeably excluded by the present experimental bounds. 
Planned searches for $d_\mu$ would also get close to test the range 
corresponding to radiatively induced misalignments.

Let us consider in turn the predictions for the imaginary part of the products of $\d$'s, eq. (\ref{FV}),
from their expressions given in eq. (\ref{eq_dsu5}). 
As before, the most important contribution comes out from the LL RR double insertion. 
Since $\eta_{\ell} \approx \i3$ and the phase of $\mu$ - if any -  is experimentally small enough to 
be safely neglected in the present discussion, 
one has at the lowest relevant order in the $t$'s
\beq
 {\cal I}m(\d^{LL} m_\ell  \d^{RR})_{ii}  = ~ 3~ t_T t_3~  \frac{(6 \m02 + 2 \A02 )^2 }{m_R^2 m_L^2}~ 
 {\cal I}m( {\cal N}  m_\ell U^* )_{ii}   ~~.
\label{su5_LLRR}
\eeq
This FV contribution is potentially much bigger than the FC one because, apart from Yukawas
and numerical coefficients, it is enhanced by a factor $(m_\tau \mu \tgb) /(m_i a_0)$.

The other FV combinations in (\ref{FV}) are:
\bea
 {\cal I}m(\d^{LR} m_\ell \d^{LR})_{ii} &=&
  ~12 ~t_T t_3 ~  \frac{ \A02  }{m_R^2}~  \frac{   {\cal I}m( {\cal N} m_\ell^3 U^*  )_{ii }}{m_L^2 }
\label{su5_LRLR} \\
 {\cal I}m(\d^{LR} \d^{RR})_{ii}& = & ~ 6~ t_T t_3~ \frac{a_0 (6 \m02 + 2 \A02 ) }{m_R^3}~ 
 \frac{{\cal I}m({\cal N}m_\ell U^*  )_{ii}}{m_L} ~~ \label{su5_LRRR} \\
 {\cal I}m(\d^{LL} \d^{LR})_{ii}& = & \frac{m_R^2}{m_L^2} {\cal I}m(\d^{LR} \d^{RR})_{ii} ~~. \label{su5_LLLR}
\eea
The only contribution which does not vanish in the limit $a_0 \rightarrow 0$ is the LL RR one. 
To compare the amplitudes,  (\ref{su5_LLRR}), (\ref{su5_LRLR})  and (\ref{su5_LRRR}), (\ref{su5_LLLR}) 
have to be multiplied respectively by $\mu \tgb$ and $m_L m_R$ and also
by the appropriate loop functions, 
which have the same sign and in general are of the same order of magnitude (for more details see \cite{ms1}).
Then, if $\mu > 0$  the four FV amplitudes have the same sign. 
However, the double LR insertion is always negligible with respect to the LL RR one because of the 
suppression factor $m_\ell^2 / m_L^2$. 
For the other amplitudes (\ref{su5_LRRR}),(\ref{su5_LLLR}) the suppression factor 
with respect to (\ref{su5_LLRR}) is $m_{R,L}/(\mu \tgb)$ (actually smaller due to the numerical coefficients). 
Then, even in the case of $\mu < 0$, a reduction of the LL RR amplitude due to accidental cancellations
seems unrealistic.
Of course, also the contributions due to different thresholds in the right-handed neutrino spectrum are 
present, in exact analogy to what has been discussed in the previous section. 

It is instructive to focus on the magnitude and dependencies of the combination 
${\cal I}m({\cal N}   m_\ell U^* )_{ii}$. 
For $d_e$ and $d_\mu$, neglecting subleading terms proportional to $y_c^2$, $y_u^2$
and defining $V_{td} \equiv |V_{td}| e^{i \phi_{td}}$: 
\bea
{\cal I}m({\cal N}  m_\ell U^*)_{22} 
                 &  \approx &  m_\tau~ y_t^2~ V_{ts}~ \left(   {\cal I}m({\cal N}_{23})   
                                           -  \frac{m_e}{m_\tau} ~ |V_{td}| ~{\cal I}m( e^{i \phi_{td}} {\cal N}_{12})\right) \\
{\cal I}m({\cal N}  m_\ell U^*)_{11} 
                       &    \approx & m_\tau ~ y_t^2  |V_{td}| \left( {\cal I}m(e^{-i \phi_{td}}  {\cal N}_{13}) 
                                   +  \frac{m_\mu}{m_\tau} ~ V_{ts} ~{\cal I}m( e^{-i \phi_{td}} {\cal N}_{12}) \right) 
\eea
where we exploited the fact that $V_{ts}$ is real in the standard parameterization.  
The latter is convenient to stress that the CP phases involved in the above combinations
could be naturally large - as is indeed the case for $\phi_{td}$ - but, of course, 
any other choice must give equivalent results
\footnote{Had we exploited the freedom of parameterizing $V$ in such a way that 
$V_{ti}$ are real numbers, then the CKM phase of $V$ would have been hidden in the redefinition of ${\cal N}$.}.
The contribution proportional to ${\cal I}m({\cal N}_{12})$ 
has important suppression factors. Moreover $|{\cal N}_{12}|$ is independently constrained to be quite small
from the present limits on $\meg$. A plot of the present upper limit on $|C_{12}|$,
with $C \equiv (4 \pi)^2 t_3 {\cal N}$, in the plane $(\ti M_1, m_R)$ can be found in Ref. \cite{imsusy02}.
The limit were derived for the seesaw but also applies without significant 
modifications to the case of $SU(5)$ plus seesaw. 
As a result, once fixed $M_T$, 
experimental searches for $d_e$ and $d_\mu$ represent a test for 
${\cal I}m(e^{-i \phi_{td}} {\cal N}_{13})$ and ${\cal I}m({\cal N}_{23})$ respectively.
Although present searches for $\tmg$ ($\tau \rightarrow e \gamma$)
are not able by now to interestingly constrain 
$|C_{23}|$ ($|C_{13}|$), eventual experimental improvements would have an impact on
$d_\mu$ ($d_e$) \cite{inprog}.

Notice also that the naive scaling relation is violated according to
\beq
\frac{d_e}{d_\mu} = \frac{|V_{td}|}{V_{ts}} \frac{{\cal I}m(e^{-i \phi_{td}} {\cal N}_{13})}{{\cal I}m({\cal N}_{23})}
\label{scsu5}
\eeq
and that the combinations of Yukawas relevant for $d_i$ 
are independent on the phases of the diagonal $SU(3)$ matrix, 
$\phi_u$. On the contrary, the latter affects (see for instance Ref. \cite{gotonihei}) 
the proton decay lifetime due to d=5 operators, whose 
most important decay mode in the case of minimal $SU(5)$ is $p \rightarrow K^+ \bar \nu$. 

\subsection{Predicted range and constraints on Yukawas}

\begin{figure}[!h]
\vskip .5 cm
\centerline{
\psfig{file=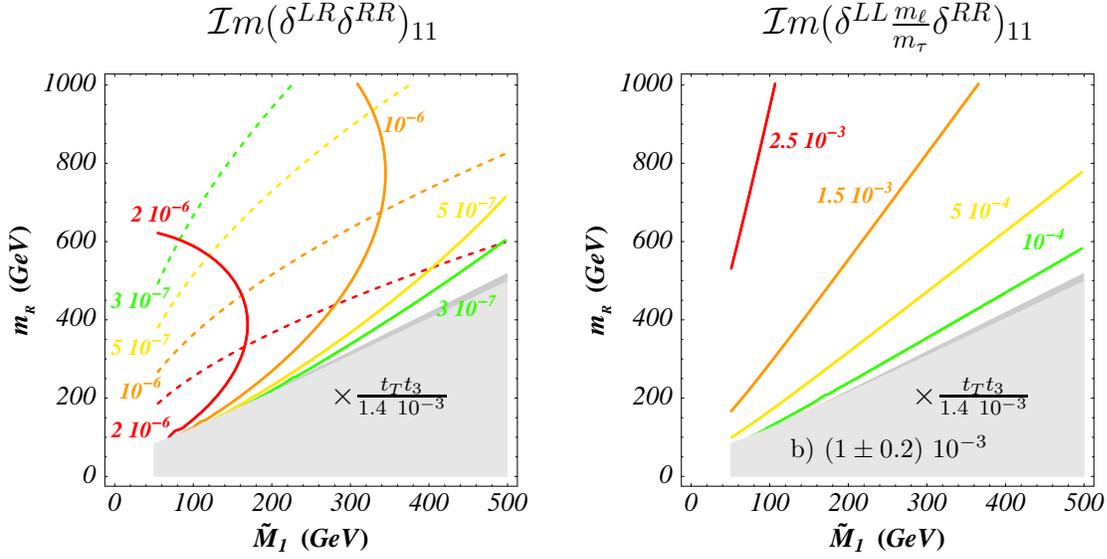,width=1. \textwidth}
\put(-320,230){ \Large Theory: $SU(5)$ with Seesaw } 
\put(-350,200){ \large ${\cal I}m(\d^{LR} \d^{RR})_{11}$} 
\put(-140,200){ \large ${\cal I}m(\d^{LL} \frac{m_\ell}{m_\tau} \d^{RR})_{11}$} 
\put(-300,60){$\times \frac{t_T t_3}{1.4~ 10^{-3}} $ } 
\put(-130,40){ \footnotesize b) $  (1 \pm 0.2)~10^{-3} $ }
\put(-80,60){$\times \frac{t_T t_3}{1.4~ 10^{-3}}  $ } }
\caption{Upper estimate for various products of $\d$'s in $SU(5)$ with seesaw.
The reference values $\Lambda /M_T =10^2$ and $\Lambda / M_3 = 2 \cdot 10^3$ have been taken. 
Solid (dashed) lines correspond to case a) (b)).  }
\label{Fsu5_de}
\end{figure}

\begin{figure}[!h]
\vskip 1.3 cm
\centerline{
\psfig{file=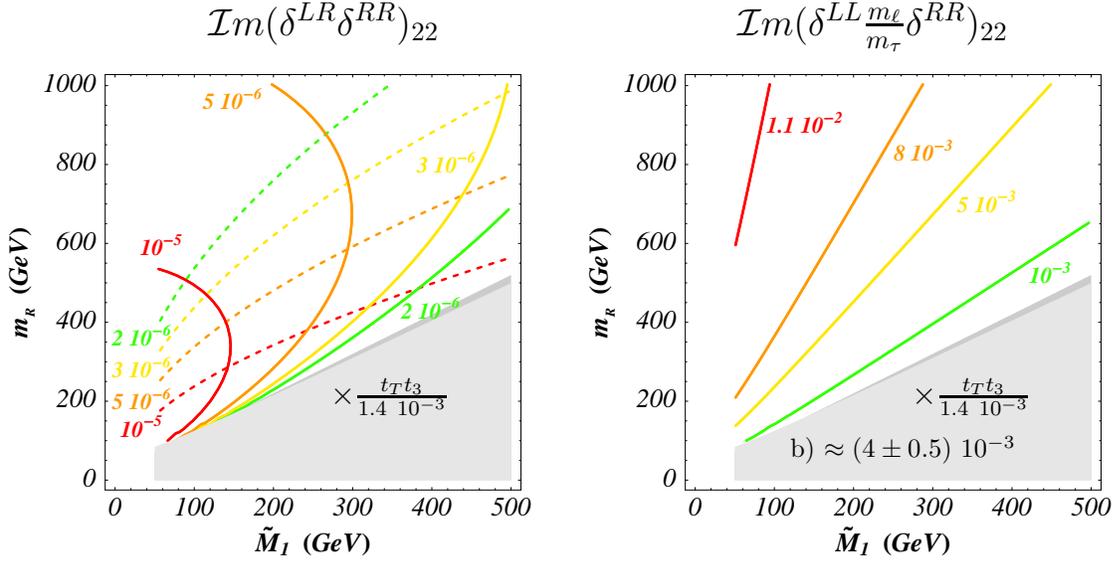,width= 1 \textwidth}
\put(-300,230){\Large Theory: $SU(5)$ with Seesaw }
\put(-350,200){ \large ${\cal I}m(\d^{LR} \d^{RR})_{22}$} 
\put(-150,200){ \large ${\cal I}m(\d^{LL} \frac{m_\ell}{m_\tau} \d^{RR})_{22}$}
\put(-300,60){$\times \frac{t_T t_3}{1.4~ 10^{-3}} $ } 
\put(-130,40){ \footnotesize b) $ \approx (4 \pm 0.5)~10^{-3} $ } 
\put(-80,60){$\times \frac{t_T t_3}{1.4~ 10^{-3}}  $ }  }
\caption{Same as fig. \ref{Fsu5_de} but for $i=2$.}
\label{Fsu5_dmu}
\end{figure}

To understand how close to the experimental sensitivity is the radiatively induced EDM range, 
we plot in figs. \ref{Fsu5_de} and \ref{Fsu5_dmu} the upper estimate for the most important products of 
$\d$'s. We consider a degenerate spectrum of right-handed neutrinos to pick out just the effect of the triplets,
substitute the $M_{Pl}$-values for $y_t$ ($\approx .7$) and the relevant CKM elements and choose as
reference values $\Lambda / M_T =10^2$ and $\Lambda / M_3 = 2~ 10^3$.   
Then, the upper estimate follows by requiring perturbativity, 
${\cal I}m(N_{23}) \le 1$, ${\cal I}m(e^{-i \phi_{td}} N_{13}) \le 1$.
Solid and dashed lines refers to cases a) and b) respectively. 
We don't show case b) for ${\cal I}m(\d^{LL} m_\ell \d^{RR})_{ii}/m_\tau$, because the predicted 
value is essentially flat.
The upper estimate for ${\cal I}m(\d^{LL} \d^{LR})_{ii}$ is not shown, being closely related to that on 
${\cal I}m(\d^{LR} \d^{RR})_{ii}$ (see eq. (\ref{su5_LLLR})). 
The bounds on ${\cal I}m(\d^{LR} m_\ell \d^{LR})_{ii}/m_\tau$ are not displayed  because they are too 
small to be of potential interest.  

\begin{figure}[!h]
\vskip .3 cm
\centerline{\LARGE Experiment} \vskip .2 cm
\centerline{
\psfig{file=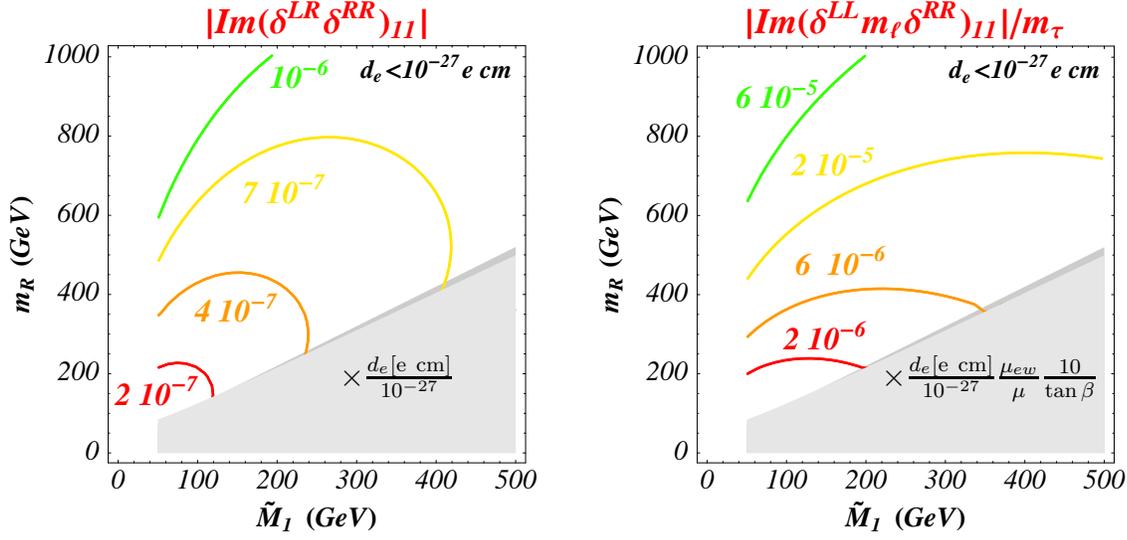,width=1.05 \textwidth}
\put(-320,60){$\times \frac{d_e [{\rm e~cm}]}{10^{-27}} $}
\put(-115,60){$\times \frac{d_e [{\rm e~cm}]}{10^{-27}} \frac{\mu_{ew}}{\mu}  \frac{10}{\tgb} $}}
\caption{Experimental upper bounds \cite{ms1} on various products of $\d$'s 
corresponding to the present sensitivity $d_e < 10^{-27}$ e cm.}
\label{Fedm_de}
\end{figure}

\begin{figure}[!h]
\vskip .3 cm
\centerline{\LARGE Experiment} \vskip .2 cm
\centerline{\psfig{file=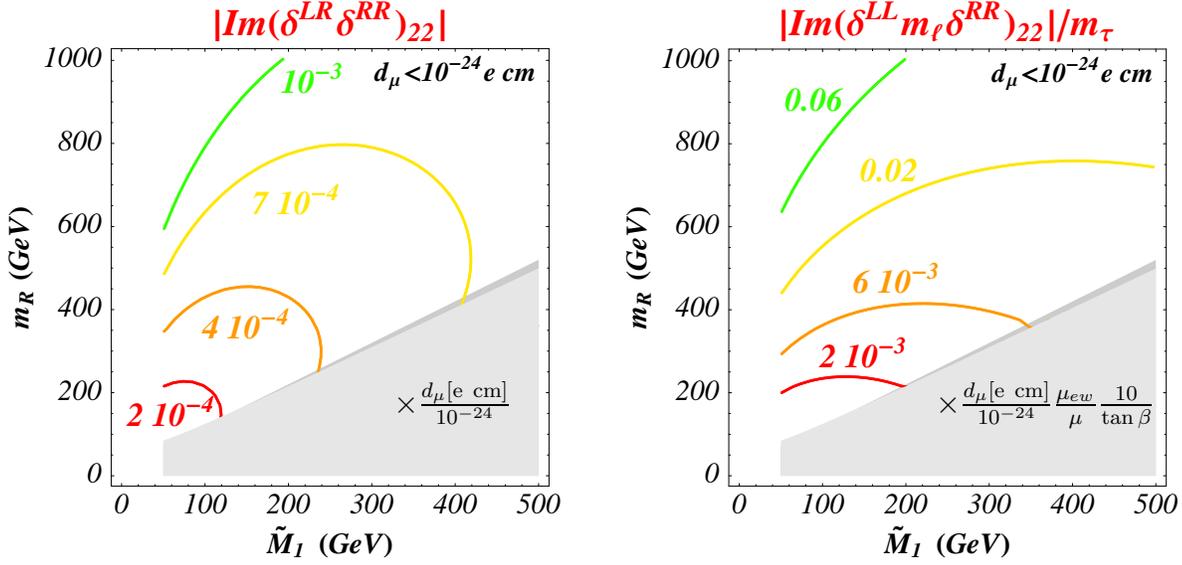,width=1.1 \textwidth}
\put(-320,60){$\times \frac{d_\mu [{\rm e~cm}]}{10^{-24}} $}
\put(-115,60){$\times \frac{d_\mu [{\rm e~cm}]}{10^{-24}}  \frac{\mu_{ew}}{\mu}  \frac{10}{\tgb}$}}
\caption{Same as fig. \ref{Fedm_de} but for $d_\mu < 10^{-24}$ e cm.}
\label{Fedm_dmu}
\end{figure}

For an easy comparison, we have taken from the sleptonarium \cite{ms1} the experimental limits 
on the same quantities, 
figs. \ref{Fedm_de} and \ref{Fedm_dmu} .
The experimental limits on ${\cal I}m(\d^{LR} \d^{RR})_{ii}$ are close to those on  
${\cal I}m(\d^{LL} \d^{LR})_{ii}$ and mildly depend on $m_L$, which has been fixed as in mSUGRA in the plots.
On the contrary, the limits on ${\cal I}m(\d^{LL} m_\ell \d^{RR})_{ii}/m_\tau$ 
are proportional to $1/ (\mu \tan \beta)$.  
For definiteness, $\tan \beta =10$ and $\mu=\mu_{ew}$ have been assumed.
The experimental bounds shown in figs. \ref{Fedm_de} and \ref{Fedm_dmu} correspond to the present 
bound $d_e < 10^{-27}$ e cm and to the planned sensitivity $d_\mu < 10^{-24}$ e cm. 
Since the experimental bounds are proportional to the bound on $d_i$,
it is straightforward to extrapolate the sensitivity to $d_\mu$ and $d_e$ required to test the radiatively induced 
$d_e$ and $d_\mu$. 

\begin{figure}[!h]
\centerline{
\psfig{file=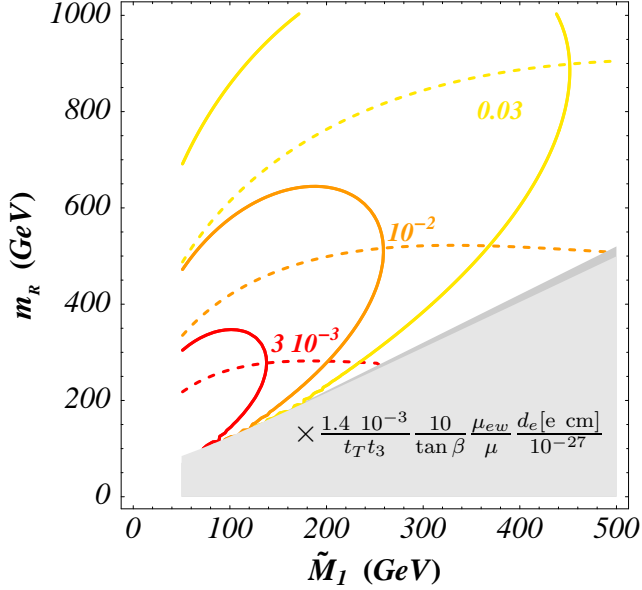,width=.6 \textwidth}
\put(-155,270){\large Upper limit on}
\put(-150,240){\large ${\cal I}m(e^{-i \phi_{td}} {\cal N}_{13})$}
\put(-145,60){$\times \frac{1.4~ 10^{-3}}{t_T t_3} \frac{10}{\tgb} \frac{\mu_{ew}}{\mu} 
\frac{d_e [{\rm e~ cm}]}{10^{-27}}$}  }
\caption{Present upper bound on ${\cal I}m(e^{-i \phi_{td}} {\cal N}_{13})$. The reference values 
$\tgb = 10$, $\Lambda /M_T =10^2$ and $\Lambda / M_3 = 2 \cdot 10^3$ have been taken.
Solid (dashed) lines correspond to case a) (b)).}
\label{Fsu5_N13}
\end{figure}

Let us firstly discuss the $d_e$ range.
For our reference values, the upper estimate for ${\cal I}m(\d^{LR} \d^{RR})_{11}$  
is $O(10^{-6})$ and the present bound on $d_e$ already constrains it to be smaller 
in a large region of the plane.  
Although we already know that they are not dominant,
it is worth to discuss the LR RR amplitude and the similar LL LR one 
because, as already mentioned, they are quite model independent. 
Allowing for higher triplet masses, however, these amplitudes shift below the present experimental sensitivity.
This is not the case for the LL RR amplitude. 
The maximum value allowed for ${\cal I}m(\d^{LL} m_\ell \d^{RR})_{11}/m_\tau$ is displayed in the right
panel of fig. \ref{Fsu5_de} for case a). In case b) it is $\approx 10^{-3}$ everywhere.
Then, the present experimental bound on $d_e$ has already explored the radiative range
for roughly $2-3$ orders of magnitude.

This can be translated into an upper limit on ${\cal I}m(e^{-i \phi_{td}} {\cal N}_{13})$, as in fig. \ref{Fsu5_N13}.
Notice that this limit is indeed very strong and can be hardly evaded even looking for less favorable 
parameters than those taken as reference. 
To check this, keep e.g. $t_3$ fixed and try to worsen the limit:
a reduction of $\mu$ with respect to $\mu_{ew}$ is unlikely to reduce it more than one 
order of magnitude;
a suppression by another factor $10$ would require a $\Lambda / M_T \approx 1.6$ ;
on the other hand, for values of $\tgb$ larger than 10 the limit linearly improves.
As a result, in the framework of the seesaw accompanied by $SU(5)$, 
the limit on ${\cal I}m(e^{-i \phi_{td}} {\cal N}_{13})$ can be considered robust.

\begin{figure}[!h]
\centerline{
\psfig{file=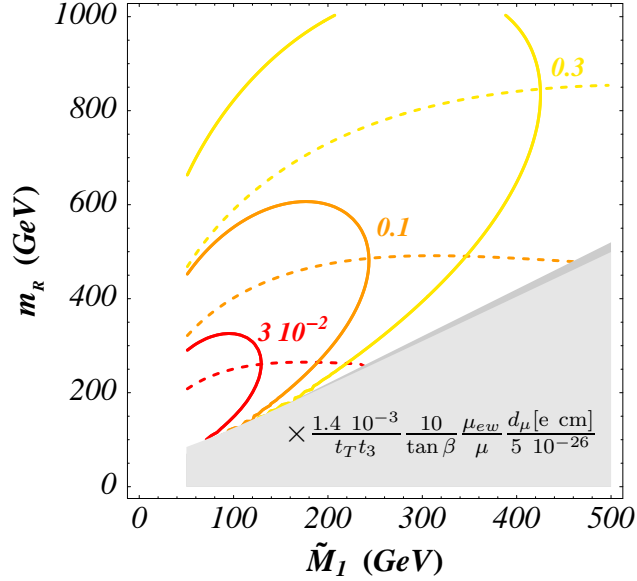,width=.6 \textwidth}
\put(-155,270){\large Upper limit on}
\put(-140,240){\large ${\cal I}m({\cal N}_{23})$}
\put(-150,60){$\times \frac{1.4~ 10^{-3}}{t_T t_3} \frac{10}{\tgb} \frac{\mu_{ew}}{\mu} 
\frac{d_\mu [{\rm e~ cm}]}{5~10^{-26}}$}}
\caption{The upper bound on ${\cal I}m({\cal N}_{23})$ which would be extracted by
improving the present limit by many orders of magnitude \cite{dmuexpf, dmuexpff}. 
Solid (dashed) lines correspond to case a) (b)). }
\label{Fsu5_N23} \vskip 1 cm
\end{figure}

Let us now turn to discuss $d_\mu$. 
By comparing figs. \ref{Fedm_dmu} and \ref{Fsu5_dmu}, it turns out that 
the upper bound on the LR RR insertion would require a sensitivity to $d_\mu$ at $O(10^{-26})$ e cm, 
at the very limit of planned experimental improvements. 
Quantitatively, the upper estimate for 
${\cal I}m(\d^{LL} m_\ell \d^{RR})_{22}/m_\tau$ which, for case b) is $\approx 4 ~10^{-3}$, is more promising. 
The major part of the plane in fig. \ref{Fsu5_dmu} could be tested with $d_\mu$ at the level of 
$10^{-24}$--$10^{-25}$ e cm and in general $d_\mu$ should not exceed $O(10^{-23})$ e cm.
As a result, the eventual presence of triplets doesn't enhance by much the range for $d_\mu$ with
respect to the pure seesaw case. 
Nevertheless, here too the possibility of constraining ${\cal I}m({\cal N}_{23})$ can be envisaged, as 
shown in fig. \ref{Fsu5_N23}. Notice that, due to (\ref{scsu5}), a limit on ${\cal I}m({\cal N}_{23})$
comparable to the present one on  ${\cal I}m(e^{-i \phi_{td}} {\cal N}_{13})$ would require
to improve the $d_\mu$ sensitivity down to $5~ 10^{-27}$ e cm.

\vskip 1 cm


\section{Conclusions}

Planned experiments might significantly strengthen the limit on $d_e$ \cite{deexpf, lam} and $d_\mu$
\cite{dmuexpf, dmuexpff}. 
Their eventual discovery could be interpreted as an indirect manifestation of supersymmetry 
but could not reveal {\it which source} of CP (and possibly flavour) violation is actually 
responsible for the measured effect. 
Clearly, all sources in principle able to give the lepton
EDM even at an higher level would be automatically constrained
while those which fail in giving the lepton EDM at the desired level would be automatically 
excluded from the list of possible candidates. 

In this work, we have estimated the ranges for the lepton EDMs induced by 
the Yukawa interactions of the heavy neutrinos, both alone and with the simultaneous presence
of the heavy $SU(5)$ triplets.
It turns out that the FV LLRR amplitude is in general larger or comparable to the FC one.
So, EDMs are enhanced for large values of $\tan \beta$ and do not strongly depend on $a_0$. 

The pure seesaw, even with large $\tgb$ and very hierarchical right-handed neutrinos, 
cannot account for $d_\mu$ above $10^{-23}$ e cm.
Its eventual discovery above this level would then signal the presence of some  
source of CP and LF violation other than the neutrino Yukawa couplings. 
The heavy triplets Yukawa couplings would be excluded from the list of possible sources
because their additional presence do not significantly enhance the predicted
range for $d_\mu$. Notice however that in the latter case a hierarchical 
right-handed neutrino spectrum is no more essential to end up with $d_\mu$ 
at an interesting level for planned searches.
From the theoretical point of view, finding $d_\mu$ above $10^{-23}$ e cm would indeed have a remarkable 
impact.

Interestingly enough, the present experimental sensitivity to $d_e$ 
is already testing the simultaneous presence of triplets and right-handed neutrinos. 
Correspondingly, 
constrains on ${\cal I}m(e^{- i \phi_{td}} {\cal N}_{13})$ have been derived which are significant 
even for quite large values of the triplet mass and unfavorable supersymmetric masses. 
Without the triplets, the radiatively-induced $d_e$ is close to
the present experimental sensitivity only in models with large $\tgb$ and small slepton masses.
Therefore, an experimental improvement would eventually provide interesting limits  
on the imaginary part of the relevant combination of neutrino Yukawa 
couplings and right-handed neutrino masses. 

In the present discussion we have been looking for results as general as possible. 
Indeed, the specification of any particular seesaw model has been avoided 
and the attention has rather focused on the dependencies on the supersymmetric 
masses and heavy thresholds. 
Although some relevant seesaw models deserve a dedicated analysis \cite{inprog},
figs. \ref{Fss_INN}, \ref{Fsu5_N13} and \ref{Fsu5_N23} are suitable for a quick check 
of the status of any given seesaw model with respect to the present and planned experimental
limits on lepton EDMs.


\section*{Acknowledgements}

We thank C.A. Savoy for useful discussions and collaboration in the early stage of this work.
I.M. acknowledge the CNRS and the ``A. Della Riccia'' Foundation for support and the SPhT, CEA-Saclay, 
for kind hospitality. 
We also acknowledge M. Peskin and Y. Farzan for pointing out a wrong numerical coefficient 
in the previous version of eq. (\ref{FR_ss}); the impact of this correction is quantitatively 
negligible and the subsequent analysis and results stay unaffected.

\vskip 2 cm

\appendix


\footnotesize

\section{Appendix: Seesaw}

In the basis where charged fermion and right handed Majorana neutrino masses are diagonal
\beq
{W} \ni u^{cT} y_u Q H_u + d^{cT} y_d Q H_d 
            + e^{cT} y_e L H_d + \nu^{cT} y_\nu L H_u +\frac{1}{2} \nu^{cT} M_R \nu^c
\label{W}
\eeq
where $Q=(u ~d)^T$, $L=(\nu ~e)^T$ and $\langle H_{d(u)}^0 \rangle = v_{d(u)}$. 
Soft scalar masses are defined as
\bea
{\cal L}_{soft} & \ni &\ti u_R^\dagger m^2_u \ti u_R + \ti d_R^\dagger m^2_d \ti d_R +\ti Q^\dagger m^2_Q \ti Q +
                    \ti e_R^\dagger m^2_e \ti e_R +\ti L^\dagger m^2_L \ti L +\ti\nu_R^\dagger m^2_\nu \ti\nu_R 
\nn\\
                 & +& (\ti u_R^\dagger A_u \ti u_L v_u + \ti d_R^\dagger A_d \ti d_L v_d +
                     \ti e_R^\dagger A_e \ti e_L v_d + \ti \nu_R^\dagger A_\nu \ti \nu_L v_u +h.c.)
\eea
Let us also introduce the following notations:
\bea
y_x^\dagger y_x \equiv X \quad y_x y_x^\dagger \equiv \ti X \quad~~~~~~~~~~~~~~~~ (x=e,u,d) \nn\\
P_a y_\nu \equiv y_\nu^{(a)} \quad   y_\nu^\dagger P_a y_\nu \equiv \a \quad 
P_a y_\nu y_\nu^\dagger P_a  \equiv \ti \a \quad ~~~~(a=1,2,3) \nn
\eea
where $P_2$, $P_1$ project out $M_3$ and $M_{3,2}$ respectively, 
$P_2={\rm diag}(1,1,0) $, $ P_1={\rm diag}(1,0,0)$, and $P_3 = \i3$.

\subsection{Running}

Defining $t \equiv \frac{1}{(4 \pi)^2} \ln Q$, 
the running of the Yukawa coupling constants is governed by:
\bea
\frac{d y_\nu^{(a)}}{dt} &=& y_\nu^{(a)} [3 \a + E + D_\nu^{(a)}]  \nn\\
\frac{d y_e}{dt} &=& y_e [3 E + \a + D_e]   \nn \\
\frac{d y_u}{dt} &=& y_u [3 U + D + D_u^{(a)}]   \\
\frac{d y_d}{dt} &=& y_d [3 D + U + D_d]  \nn\\
\frac{d (P_a M_R P_a) } {dt} &=& 2   [ P_a M_R  \ti \a^T + \ti \a M_R P_a] \nn
\eea
where 
\bea
D_\nu^{(a)}& =& [Tr(3 U + \a ) - (3 g_2^2 + \frac{3}{5} g_1^2)] \i3 \nn\\
D_e &=& [Tr(3 D + E ) - (3 g_2^2 + \frac{9}{5} g_1^2)] \i3 \nn\\
D_u^{(a)}& =& [Tr(3 U + \a ) - (\frac{16}{3} g_3^2 +3 g_2^2 + \frac{13}{15} g_1^2 )] \i3 \\
D_d &=& [Tr(3 D + E ) - (\frac{16}{3} g_3^2 +3 g_2^2 + \frac{7}{15} g_1^2)] \i3 ~.\nn
\eea
For the trilinear couplings, defining $P_a A_\nu \equiv A_\nu^{(a)}$ 
\bea
\frac{d A_\nu^{(a)}}{dt} &=& 4 \ti \a  A_\nu^{(a)} + 5 A_\nu^{(a)} \a +2 y_\nu^{(a)} y_e^\dagger A_e + A_\nu^{(a)} E + 
D_\nu^{(a)} A_\nu^{(a)} + 2 \ti D_\nu^{(a)} y_\nu^{(a)}
\nn\\
\frac{d A_e}{dt} &=& 4 \ti E A_e + 5 A_e E +2 y_e {y_\nu^{(a)} }^\dagger A_\nu^{(a)} +A_e \a + D_e A_e + 2 \ti D_e y_e 
\label{Ae_ss}
\\
\frac{d A_u}{dt} &=& 4 \ti U A_u + 5 A_u U +2 y_u y_d^\dagger A_d +A_u D + D_u^{(a)} A_u + 2 \ti D_u^{(a)} y_u
\nn\\
\frac{d A_d}{dt} &=& 4 \ti D A_d + 5 A_d D +2 y_d y_u^\dagger A_u +A_d U + D_d A_d + 2 \ti D_d y_d \nn
\eea
where
\bea
\ti D_\nu^{(a)} & =& [Tr(3 y_u^\dagger A_u+ {y_\nu^{(a)}}^\dagger A_\nu^{(a)} ) - (3 g_2^2 \ti M_2  
+ \frac{3}{5} g_1^2 \ti M_1)] \i3 \nn\\
\ti D_e &=& [Tr(3 y_d^\dagger A_d+ y_e^\dagger A_e ) - (3 g_2^2 \ti M_2  + \frac{9}{5} g_1^2 \ti M_1)] \i3 \\
\ti D_u^{(a)} &=& [Tr(3 y_u^\dagger A_u+ {y_\nu^{(a)}}^\dagger A_\nu^{(a)} ) - (\frac{16}{3}g_3^2 \ti M_3 
+ 3 g_2^2 \ti M_2  + \frac{13}{15}
 g_1^2 \ti M_1)] \i3 \nn\\
\ti D_d & =& [Tr(3 y_d^\dagger A_d+ y_e^\dagger A_e ) - (\frac{16}{3}g_3^2 \ti M_3+3 g_2^2 \ti M_2  + \frac{7}{15} 
g_1^2 \ti M_1)] \i3 ~.\nn
\eea
For soft scalars, defining $P_a m^2_\nu P_a \equiv {m^2_\nu}^{(a)}$
\bea
\frac{d m^2_L}{dt} &=& \{ m^2_L , E + \a \} +2 ( y_e^\dagger m^2_e y_e + m^2_{H_d} E + A_e^\dagger A_e) 
+ 2 ( {y_\nu^{(a)}}^\dagger {m^2_\nu}^{(a)} y_\nu^{(a)} + m^2_{H_u} \a + {A_\nu^{(a)}}^\dagger A_\nu^{(a)}) 
+ G_L \nn\\
\frac{d m^2_e}{dt} &=& 2  \{ m^2_e , \ti E  \} +4 ( y_e  m^2_L  y_e^\dagger + m^2_{H_d} \ti E + A_e A_e^\dagger ) 
+ G_e \nn\\
\frac{d {m^2_\nu}^{(a)} }{dt} &=& 2  \{ {m^2_\nu}^{(a)} , \ti \a  \} +4 ( y_\nu^{(a)}  m^2_L  {y_\nu^{(a)}}^\dagger + 
m^2_{H_u} \ti \a + 
A_\nu^{(a)} {A_\nu^{(a)}}^\dagger )  \\
\frac{d m^2_Q}{dt} &=& \{ m^2_Q , U + D \} +2 ( y_u^\dagger m^2_u y_u + m^2_{H_u} U + A_u^\dagger A_u) 
+ 2 ( y_d^\dagger m^2_d y_d + m^2_{H_d} D + A_d^\dagger A_d) + G_Q \nn\\
\frac{d m^2_u}{dt} &=& 2  \{ m^2_u , \ti U  \} +4 ( y_u  m^2_Q  y_u^\dagger + m^2_{H_u} \ti U + A_u A_u^\dagger ) 
+ G_u \nn\\
\frac{d m^2_d}{dt} &=& 2  \{ m^2_d , \ti D  \} +4 ( y_d  m^2_Q  y_d^\dagger + m^2_{H_d} \ti D + A_d A_d^\dagger ) 
+ G_d \nn
\eea
where
\bea
G_L = -(\frac{6}{5} g_1^2 \ti M_1^2 +6 g_2^2 \ti M_2^2) \i3 \quad~~\quad~~~~~~~~~~~~~~~~
G_e = -(\frac{24}{5} g_1^2 \ti M_1^2) \i3  \nn\\
G_Q = -(\frac{2}{15} g_1^2 \ti M_1^2 +6 g_2^2 \ti M_2^2 +\frac{32}{3} g_3^2 \ti M_3^2) \i3 \quad~~~~~ \\
G_u = -(\frac{32}{15} g_1^2 \ti M_1^2 +\frac{32}{3} g_3^2 \ti M_3^2) \i3 \quad~~\quad~~~~
G_d = -(\frac{8}{15} g_1^2 \ti M_1^2 +\frac{32}{3} g_3^2 \ti M_3^2) \i3 ~.\nn
\eea
Finally,
\bea
\frac{d m^2_{H_d}}{dt} &=& 6 Tr(y_d m^2_Q y_d^\dagger +y_d^\dagger m^2_d y_d + m^2_{H_d} D + A_d^\dagger A_d)
+ 2 Tr(y_e m^2_L y_e^\dagger +y_e^\dagger m^2_e y_e + m^2_{H_d} E + A_e^\dagger A_e) + G_L \nn\\
\frac{d m^2_{H_u}}{dt} &=& 6 Tr(y_u m^2_Q y_u^\dagger +y_u^\dagger m^2_u y_u + m^2_{H_u} U + A_u^\dagger A_u)
\\ &+& 2 Tr(y_\nu^{(a)} m^2_L {y_\nu^{(a)}}^\dagger +{y_\nu^{(a)}}^\dagger {m^2_\nu}^{(a)} y_\nu^{(a)} + m^2_{H_u} \a 
+ {A_\nu^{(a)}}^\dagger A_\nu^{(a)}) + G_L \nn
\eea
and we define $ {m^2}'_{H_{d(u)}} \equiv d m^2_{H_{d(u)}} /  dt $.

For energy scales between $\Lambda$ and $M_3$, one has to take $a=3$;
below $M_3$ and above $M_2$, $a=2$; while  
below $M_2$ and above $M_1$, $a=1$.


\section{Appendix: $SU(5)$ + See-saw}

We adopt the following notation to write matter and Higgs superfields:
\beq
\psi_{10}= \frac{1}{\sqrt{2}} \left( \matrix{0 & u_3^c  & - u_2^c & u^1 &  d^1 \cr
                                                       - u_3^c & 0          &  u_1^c  & u^2 &  d^2 \cr
                                                         u_2^c &   -u_1^c &   0       & u^3  &  d^3 \cr
                                                      - u^1  &    - u^2      &  -u^3    &    0   &  e^c \cr
                                                     - d^1     & - d^2      &  -d^3    &    - e^c & 0     }   \right)  \quad
\phi_{\bar 5} = \left(  \matrix{d_1^c \cr d_2^c \cr d_3^c \cr e \cr -\nu} \right)  \quad \eta_1 = \nu^c
\eeq
\beq
\bar H =  \left(  \matrix{{H_{3d}}_1 \cr {H_{3d}}_2 \cr {H_{3d}}_3 \cr {H_{2d}}^- \cr - {H_{2d}}^0} \right)  \quad
H =  \left(  \matrix{{H_{3u}}^1 \cr {H_{3u}}^2 \cr {H_{3u}}^3 \cr {H_{2u}}^+ \cr  {H_{2u}}^0} \right)  \quad
\eeq
where $< H_{2d(2u)}^0 > \equiv  v_{d(u)}$. 
The superpotential 
\beq
W \ni \frac{1}{4} ~{\psi^{AB}} y_u ~\psi^{CD} H^E \epsilon_{ABCDE} + \sqrt{2}~  {\psi^{AB}} y_e ~ \phi_A \bar H_B 
       + \eta_1 y_\nu ~\phi_A H^A + \frac{1}{2} \eta_1 M_R~ \eta_1
\eeq
where $A,B, ... =1,...,5$ and flavour indices are understood, gives rise to (\ref{W})
with $y_d=y_e^T$ and $y_u=y_u^T$. 
For the soft breaking part of the Lagrangian 
\bea
{\cal L}_{soft}& =& {\ti \psi}^\dagger m^2_\psi {\ti \psi} + \ti \phi^\dagger m^2_\phi \ti \phi +
\ti \eta^\dagger m^2_\eta \ti \eta + m^2_h h^\dagger h + m^2_{\bar h} {\bar h}^\dagger {\bar h} +
( \frac{1}{2} M_5 \lambda_5 \lambda_5 + {\rm h.c.}  ) \nn\\
&+& \left( \ti u^{cT} A_u \ti u ~v_u + \ti d^{cT} A_e^T \ti d ~v_d +  \ti e^{cT} A_e \ti e ~v_d 
+ \ti \nu^{cT} A_\nu \ti \nu~ v_u  \right) + {\rm h.c.}  
\eea
where the scalar fields are $\ti \psi=(\ti u^c, \ti u, \ti d, \ti e^c)$, $\ti \phi=(\ti d^c, \ti e, \ti \nu)$ and $\ti \eta = \ti \nu^c$
and gauginos are denoted with $\lambda_5$.
In this way, in the scalar lepton mass matrix, $m^2_{LL}=m^2_\phi$, $m^2_{RR}=m^{2*}_\psi$.

\subsection{Running}

Setting $t \equiv \frac{1}{(4 \pi)^2} \ln Q$, $d g_5/dt= -3 g_5^3$, $ d M_5/dt= -6 g_5^2 M_5$ and
\bea
\frac{d y_e}{dt}&=& 6 y_e E + 3 \ti U y_e + y_e N +G_e y_e \nn\\
\frac{d y_u}{dt}&=& 6 y_u U + 2 \ti  E y_u + 2 y_u D  +G_u y_u \\
\frac{d y_\nu}{dt} &=& 6 y_\nu N + 4 y_\nu E + G_\nu y_\nu \nn
\label{rsu5_y}
\eea
where $G_e= -84/5 g_5^2 + 4 Tr(E)$, $G_u= -96/5 g_5^2 +  Tr(3 U+ N)$, $G_\nu= -48/5 g_5^2 +  Tr(3 U+N)$.
For scalar masses
\bea
\frac{d m^{2*}_\psi}{dt}&=& \{ m^{2*}_\psi, 2 \ti E+ 3 \ti U \} + 4 (y_e  m^2_\phi y_e^\dagger 
+ m^2_{\bar h} \ti E + A_e A_e^\dagger ) + 6 (y_u  m^2_\psi y_u^\dagger 
+ m^2_{ h} \ti U + A_u A_u^\dagger ) + G_\psi \nn\\
\frac{d m^{2}_\phi}{dt}&=& \{ m^{2}_\phi, 4  E+  N \} + 8 (y_e^\dagger  m^{2*}_\psi y_e 
+ m^2_{\bar h} D +  A_e^\dagger A_e ) + 2 (y_\nu^\dagger  m^{2*}_\eta y_\nu 
+ m^2_{ h}  N + A_\nu^\dagger A_\nu ) + G_\phi \nn \\
\frac{d m^{2*}_\eta}{dt}&=& 5 \{ m^{2*}_\eta , \ti N \} + 10 (y_\nu m^{2}_\phi y_\nu^\dagger + m^2_h \ti N 
+ A_\nu A_\nu^\dagger) 
\label{rsu5_m}
\eea
where $G_\psi = -144/5 g_5^2 M_5^2 \i3$,  $G_\phi = -96/5 g_5^2 M_5^2 \i3$. For trilinear couplings
\bea
\frac{d A_e}{dt}&=& 10 \ti E A_e + 3 \ti U A_e + 8 A_e E + 6 A_u y_u^\dagger y_e + A_e N 
         + 2 y_e y_\nu^\dagger A_\nu + G_e A_e +2 \ti G_e y_e \label{Ae_su5} \\
\frac{d A_u}{dt}&=& 9 \ti U A_u + 2 \ti E A_u + 9 A_u U + 4 A_e y_e^\dagger y_u + 2 A_u D 
         + 4 y_u y_d^\dagger A_d + G_u A_u +2 \ti G_u y_u \\
\frac{d A_\nu}{dt}&=& 7 \ti N A_\nu + 11  A_\nu N + 4 A_\nu E + 8  y_\nu y_e^\dagger A_e
+ G_\nu A_\nu +2 \ti G_\nu y_\nu 
\label{rsu5_A}
\eea
where $\ti G_e= -84/5 g_5^2 M_5  + 4 Tr(y_e^\dagger A_e)$, 
$\ti G_u= -96/5 g_5^2 M_5 +  Tr(3 y_u^\dagger A_u+ y_\nu^\dagger A_\nu)$, 
$\ti G_\nu= -48/5 g_5^2 M_5 +  Tr(3 y_u^\dagger A_u + y_\nu^\dagger A_\nu)$.
Finally, for scalar higgses
\bea
\frac{d m^{2}_h}{dt}&=&Tr(6 \ti U +2 \ti N ) m^2_h  + 6 Tr(y_u  m^2_\psi y_u^\dagger 
+ y_u^\dagger m^{2*}_\psi y_u + A_u A_u^\dagger ) \nn\\ 
&+& 2 Tr(y_\nu  m^2_\psi y_\nu^\dagger + y_\nu^T m^2_\eta y_\nu^* + A_\nu A_\nu^\dagger ) 
- \frac{96}{5} g_5^2 M_5^2 \\
\frac{d m^{2}_{\bar h}}{dt}&=& 8 Tr( \ti E ) m^2_{\bar h}  + 8 Tr(y_e  m^2_\phi y_e^\dagger 
+ y_e^\dagger m^{2*}_\psi y_e + A_e A_e^\dagger ) - \frac{96}{5} g_5^2 M_5^2 ~.\nn
\eea





\end{document}